\documentclass[twocolumn,amsmath,amssymb,prb,longbibliography]{revtex4-1}
\usepackage{multirow}
\usepackage{array}
\usepackage{amsmath}
\usepackage{graphicx}% Include figure files
\usepackage{dcolumn}% Align table columns on decimal point
\usepackage{bm}% bold math
\usepackage{verbatim,xcolor}
\DeclareMathAlphabet \mathbfcal{OMS}{cmsy}{b}{n}
\begin{document}

\title{Ultrafast strong-field absorption in gapped graphene}

\author{S. Azar Oliaei Motlagh}
\author{Ahmal Jawad Zafar}
\author{Aranyo Mitra}
\author{Vadym Apalkov}
\author{Mark I. Stockman}
\affiliation{Center for Nano-Optics (CeNO) and
Department of Physics and Astronomy, Georgia State
University, Atlanta, Georgia 30303, USA
}

\date{\today}
\begin{abstract}
We  study theoretically the strong-field absorption of an ultrafast optical pulse by a gapped graphene monolayer. At low field amplitudes, the absorbance in the pristine graphene  is  equal to the universal value of $2.3$ percent. Although the ultrafast optical absorption for low field amplitudes is independent of the polarization,  linear or circular, of the applied optical pulse, for high field amplitudes, the absorption strongly depends on the pulse polarization. 
For a linearly polarized pulse, the optical absorbance is saturated at the value of $\approx 1.4$ percent for the pulse's amplitude of $\geq 0.4~\mathrm{V/\AA}$, but no such saturation is observed for a circularly polarized pulse. For the gapped graphene, the absorption of a linearly polarized pulse has a weak dependence on the bandgap, while for a circularly polarized pulse, the absorption is very sensitive to the bandgap. 
%Opening a bandgap in graphene by placing in on, for example, SiC substrate strongly modify the ultrafast absorption at small field amplitudes.  
\end{abstract}
%\pacs{}
\maketitle
%\section{Introduction} 

\section{Introduction}
The progress in generation of ultrafast intense laser pulses provides powerful tools to observe the highly-nolinesr, strong-field phenomena in solids and expand the field of strong-field optics to both small time scales and high field intensities \cite{Schiffrin_at_al_Nature_2012_Current_in_Dielectric, Apalkov_Stockman_PRB_2012_Strong_Field_Reflection, Higuchi_Hommelhoff_et_al_Nature_2017_Currents_in_Graphene, Gruber_et_al_ncomms13948_2016_Ultrafast_pulses_graphene, Stockman_et_al_PhysRevB.95_2017_Crystalline_TI,Stockman_et_al_PhysRevB.98_2018_3D_TI, 
Hommelhoff_et_al_PhysRevLett.121_2018_Coherent,  Hommelhoff_et_al_1903.07558_2019_laser_pulses_graphene, sun_et_al_nnano.2011.243_2012_Ultrafast_pulses_graphene, Mashiko_et_al_Nature_Communications_2018_ultrafast_pulse_solid, Shin_et_al_IOP_Publishing_2018_ultrafast_pulse_solid,Hommelhoff_et_al_PhysRevLett.121_2018_Coherent, %Biegert_et_al_s41467_2018_Ultrafast_pulses_graphene, 
Gruber_et_al_ncomms13948_2016_Ultrafast_pulses_graphene, Higuchi_et_al_Nature_2017, Leitenstorfer_et_al_PhysRevB.92_2015_Ultrafast_Pseudospin_Dynamics_in_Graphene, Stockman_et_al_PhysRevB.98_2018_Rapid_Communication_Topological_Resonances, Sun_et_al_Chinese_Physics_B_2017_Ultrafast_pulses_TMDC, Zhang_et_al_OSA_2018_ultrafast_pulse_TMDC}. Some of such strong-field phenomena are high harmonic generations, the ultrafast ionization, the nonlinear current generations, and the nonlinear optical absorption \cite{Ghimire_et_al_Nature_Communications_2017_HHG, Reis_et_al_Nat_Phys_2017_HHG_from_2D_Crystals,
Simon_et_al_PRB_2000_Strong_Field_Fs_Ionization_of_Dielectrics,
Rosa_et_al_Optical_Materials_Express_2017_stacking_graphene_saturable_absorbers,
Kumar_et_al_APL_2009_saturable_absorption_graphene,
Gesuele_Photonics_2019_Transient_Absorption}.

Graphene is a well known two dimensional (2D) solid made of a single layer of carbon atoms. It has a honeycomb crystal structure with two inequivalent sublattices, $A$ and $B$ \cite{Geim_et_al_Nat_Mater_2007_The_rise_of_graphene,Electronic_properties_graphene_RMP_2009}. The first Brillouin zone (BZ) of graphene is a hexagon and the 
corresponding energy dispersion is gappless at two Dirac points,
$K$ and $K^\prime$, with the massless relativistic energy dispersion \cite{Butler_et_al_Acs_Nano_2013_2D_Beyond_Graphene,Novoselov_Geim_et_al_nature04233_2D_Electrons_in_Graphene,Geim_et_al_Nat_Mater_2007_The_rise_of_graphene,Electronic_properties_graphene_RMP_2009}.  
The electron states at the Dirac points are chiral. They are characterized by nontrivial 
Berry phases $\pm \pi$, which are opposite at the $K$ and $K^\prime$ points 
 \cite{Kormanyos_et_al_2d_Materials_2015_k.p_theory_for_two_dimensional,
Ye_et_al_Nature_Nanotechnology_2016_Electrical_generation_and_control, Sun_et_al_Chinese_Physics_B_2017_Ultrafast_pulses_TMDC,
Jariwala_et_al_Asc_Nano_2014_Transition_Metal}. The corresponding Berry curvature is non-zero only at the Dirac points, at which it has a $\delta$-type singularity.  
For other 2D materials with the honeycomb crystal structure but with a finite bandgap, such as monolayers of transition metal dichalcogenides (TMDCs), the Berry curvature is non-zero within finite regions near the 
$K$ and $K^\prime$ points with the maxima at these points.  
 \cite{Sun_et_al_Chinese_Physics_B_2017_Ultrafast_pulses_TMDC,
Jariwala_et_al_Asc_Nano_2014_Transition_Metal}. Thus, opening the bandgap in graphene-like materials broadens the Berry curvature in the $K$ and $K^\prime$ valleys.  Such broadening results in the effect of 
topological resonance\cite{Stockman_et_al_PRB_2019_gapped_graphene}, which occurs during ultrafast electron dynamics  and is due to the compensation of the dynamic phase and the topological phase . 

The gapped graphene has a broken inversion symmetry: the point symmetry group reduced from $D_{6h}$ for graphene to $D_{3h}$. Consequently, a bandgap opens up at the $K$ points. Previously, we have shown that, due to the existence of the bandgap, the topological resonance appears in strong fields.  Consequently, a femtosecond intense optical pulse generates a large valley polarization 
\cite{Stockman_et_al_PRB_2019_gapped_graphene}, which is not related
to the electron spin and the spin-orbit coupling (SOC). 

Also, in the gapped graphene, in addition to the field-driven longitudinal current, there is a transverse current in the direction normal to the the applied field\cite{Stockman_et_al_J.Phys.Condens.Matter_2019_current_gapped_graphene}. This transverse current  is also due to breaking of the inversion symmetry; consequently, it strongly depends on the bandgap. 

One of the important characteristics of the interaction of 
an optical pulse with solids is the absorption coefficient. For the pristine 
graphene, in the linear regime, approximately 2.3 percent of the incident light energy is absorbed. In a sharp contrast, the nonlinear absorption, which was measured for a 80-fs optical pulse, showed a saturable behavior \cite{Kumar_et_al_APL_2009_saturable_absorption_graphene,
Gesuele_Photonics_2019_Transient_Absorption,
Rosa_et_al_Optical_Materials_Express_2017_stacking_graphene_saturable_absorbers}.
For such a long pulse, the electron dynamics in the field of the pulse is incoherent, and scattering and relaxation processes are important.  
Here, we study the absorption by the gapped graphene for an ultrashot optical pulse with the duration of just a few femtoseconds. For such a short pulse, the electron dynamics is coherent, and the system exhibits new features related to the topological resonance.  To describe different types of the graphene-like materials ,we consider the model of gapped graphene with a variable bandgap. The bandgap can be opened, for example, by applying the staggered potential, which can be realized by epitaxially growing graphene on a SiC substrate.

%In this article, we investigate the nonlinear absorption of an ultrafast optical pulse in a graphene monolayer with a tunable bandgap. The ultrafast and intense optical pulse applied to graphene has a duration of a few femtoseconds and the amplitude of several tenths of Volt per Angstrom. We predict that the ultrafast absorption of graphene can be controlled not only by the characteristics of the applied pulse such as amplitude and polarization but also by the bandgap of graphene. We also predict the "bistability" in absorption of linear polarized ultrafast pulse which can be used as ultrafast switches. There are several methods to open the gap in the energy band structure of graphene. One way of opening the gap is applying the staggered potential which can be achieved by epitaxially growing graphene on SiC substrate.

\section{Model and Main Equations}
\label{Model_and_Equations}

We consider electron dynamics in the field of the pulse with 
the duration of just a few femtoseconds. 
Taking into account that electron scattering times in graphene are of the order of or longer than 10 fs -- see Refs.\ \onlinecite{Hwang_Das_Sarma_PRB_2008_Graphene_Relaxation_Time, Breusing_et_al_Ultrafast-nonequilibrium-carrier-dynamics_PRB_2011, theory_absorption_ultrafast_kinetics_graphene_PRB_2011, Ultrafast_collinear_scattering_graphene_nat_comm_2013, Gierz_Snapshots-non-equilibrium-Dirac_Nat-Material_2013, Nonequilibrium_dynamics_photoexcited_electrons_graphene_PRB_2013}, we neglect the electron collisions and assume that the electron dynamics  is coherent. Such dynamics is  described by time-dependent Schr\"odinger equation (TDSE)
%We consider two band tight binding Hamiltonian of graphene and we introduce on-site energies with opposite sign in the Hamiltonian to have a nonzero bandgap $\Delta_g$.
\begin{equation}
i\hbar \frac{{d\Psi_{\alpha\mathbf q} }}{{dt}} = { H(t)} \Psi_{\alpha\mathbf q}~,~~~ { H}(t) = { H}_0 - e{\mathbf{F}}(t){\bf{r}},
\label{Sch}
\end{equation}
%with Hamiltonian
%\begin{equation}
%{ H}(t) = { H}_0 - e{\bf{F}}(t){\bf{r}},
%\label{Ht}
%\end{equation}  
where $\mathbf F(t)$ is the electric field of the pulse, and $e$ is  electron charge. Here we assume that the electron initially (before the pulse) is in the band $\alpha$ ($\alpha=\mathrm v$ for the valence band (VB) and $\alpha=\mathrm c$ for the conduction band (CB)) with  crystal momentum $\mathbf q$. 

Field-free Hamiltonian $H_0$ is the nearest-neighbor two-band tight binding Hamiltonian of gapped graphene \cite{Lanzara_et_al_Nat_Mat_2007_Gapped_Graphene, Kjeld_et_al_PhysRevB.79.113406_2009_Gapped_Graphene_Optical_Response, Pyatkovskiy_JPCM_2008_Plasmons_in_Gapped_Graqphene}
\begin{eqnarray}
H_0=\left( {\begin{array}{cc}
   \Delta_g/2 & \gamma f(\mathbf k) \\
   \gamma f^\ast(\mathbf k) & -\Delta_g/2 \\
  \end{array} } \right) ,
\label{H0}
\end{eqnarray}
where $\Delta_g$ is the band gap,
%[doi:10.1038/nmat2003,DOI:10.1103/PhysRevB.79.113406, doi: 10.1088/0953-8984/21/2/025506]
$\gamma= -3.03$ eV is the hopping integral, and
\begin{equation}
f(\mathbf k)=\exp\Big(i\frac{ak_y}{\sqrt{3}}\Big )+2\exp\Big(-i\frac{ak_y}{2\sqrt{3}}\Big )\cos{\Big(\frac{ak_x}{2}\Big )},
\end{equation}
where $a=2.46~\mathrm{\AA}$ is the lattice constant. 

%In the absence of the electron-electron interaction and electron relaxation, this Schr\"odinger equation correctly describes many-electron dynamics where the density matrix can be computed in terms of the solutions of Eq.\ (\ref{Sch}) as
%\begin{equation}
%\hat\rho=\sum_{\alpha\mathbf q}\left|\Psi_{\alpha\mathbf q}\right\rangle n^\mathrm{(F)}_{\alpha\mathbf q}\left\langle \Psi_{\alpha\mathbf q}\right|~,
%\label{rho}
%\end{equation}
%where $n^\mathrm{(F)}_{\alpha\mathbf q}$ is the Fermi population factor.
The energies of CB and VB can be found from Hamiltonian $H_0$ and are given by the following expression 
\begin{equation}
E_{\alpha}(\mathbf k)=\pm\sqrt{\gamma ^2\left |{f(\mathbf k)}\right |^2+\Delta_g ^2/4}~,
\label{Energy}
\end{equation}
where signs $\pm$ stand for  CB ($\alpha=c$) and VB ($\alpha=v$), respectively. 
%\begin{figure}
%\begin{center}\includegraphics[width=0.47\textwidth]{Energy_gap_1eV.png}\end{center}
%  \caption{(Color online) (a) Hexagonal lattice structure of graphene with sublattices A and B. (b) The first Brillouin zone of the reciprocal lattice of graphene with two valleys $K$ and $K^\prime$. (c) Energy dispersion is shown as a function of crystal momentum for gapped graphene (1 eV)}
 % \label{fig:Energy}
%\end{figure}%

The applied electric field generates both intraband and interband electron dynamics. The intraband dynamics is described by the Bloch acceleration theorem \cite{Bloch_Z_Phys_1929_Functions_Oscillations_in_Crystals}, which 
determines the time-dependent electron wavevector, $\mathbf k({\bf{q}},t)$, 
as follows
\begin{equation}
{{\bf{k}}}({\bf{q}},t) = {\bf{q}} + \frac{e}{\hbar }\int_{ - \infty }^t {{\bf{F}}({t^\prime})d{t^\prime}}, 
\label{kvst}
\end{equation}
where ${\bf q}$ is the initial electron wavevector. 
In relation to the Bloch trajectories (\ref{kvst}), we also define 
the separatrix as a set of initial points $\mathbf q$ for which the electron trajectories pass precisely through the corresponding $K$ or $K^\prime$ points \cite{Stockman_et_al_PhysRevB.93.155434_Graphene_Circular_Interferometry}. Its parametric equation is 
\begin{equation}
\mathbf q(t)=\mathbf K-\mathbf k(0,t), \mathrm{~~or~~}  \mathbf q(t)=\mathbf K^\prime-\mathbf k(0,t),
\label{separatrix}
\end{equation}
 where $t\in (-\infty,\infty)$ is a parameter.
 
To determine the the intraband electron dynamics, we solve TDSE in terms of  the Houston functions\cite{Houston_PR_1940_Electron_Acceleration_in_Lattice}, which are adiabatic solutions for intraband dynamics,
\begin{equation}
 \Phi^\mathrm{(H)}_{\alpha {\bf q}}({\bf r},t)=\Psi^{(\alpha)}_{\bf{k}(\bf q,t)} ({\bf r})\exp\left(i\phi^{(\mathrm d)}_{\mathrm{\alpha}}({\bf q},t)+i\phi^{(\mathrm B)}_{\mathrm{\alpha}}({\bf q},t)\right),
\end{equation}
where $ \Psi^{(\alpha)}_{\bf{k}(\bf q,t)}$ are the Bloch functions. Here the dynamic phase, $\phi^\mathrm{(D)}_{\mathrm \alpha}$, and the geometrical phase, $\phi^\mathrm{(B)}_{\mathrm \alpha}$, are defined as 
\begin{eqnarray}
\phi^\mathrm{(D)}_{\alpha}(\mathbf q,t)= \frac{-1}{\hbar} \int_{-\infty}^t dt^\prime \left(E_\mathrm \alpha[\mathbf k (\mathbf q,t^\prime)]\right),
 \label{phi}
 \\ 
 \phi^\mathrm{(B)}_{\mathrm \alpha}(\mathbf q,t)= \frac{-e}{\hbar} \int_{-\infty}^t dt^\prime \mathbf F \left(\mathbfcal{A}^{\mathrm{\alpha \alpha}}[\mathbf k (\mathbf q,t^\prime)]\right).
 \label{phi}
\end{eqnarray}
In Eq.\ (\ref{phi}), $\mathbfcal{A}^{\alpha\alpha}=\left\langle \Psi^{(\alpha)}_\mathbf q  |   i\frac{\partial}{\partial\mathbf q}|\Psi^{(\alpha)}_\mathbf q   \right\rangle $ is the intraband Berry connection for band $\alpha$, which in this model can be found analytically as
\begin{eqnarray}
\mathcal{A}_{x}^{cc}(\mathbf k)&=&\frac{-a\gamma ^2}{\gamma ^2 |f(\mathbf k)|^2+(\Delta_g/2-E_c)^2}
 \sin \frac{3ak_y}{2\sqrt{3}}\sin{\frac{ak_x}{2}}
\nonumber \\
 %&& =|D_x({\mathbf k})|\exp{(i\phi_{x}({\mathbf k}))},
 \label{Axcc}
 \\
\mathcal{A}_{y}^{cc}(\mathbf k)&=&\frac{a\gamma ^2}{\sqrt{3}(\gamma ^2 |f(\mathbf k)|^2+(\Delta_g/2-E_c)^2)}\nonumber\\
 &&\times \Big(\cos{ ak_x}-\cos{\frac{\sqrt{3}ak_y}{2}}\cos{\frac{ak_x}{2}}\Big)
\nonumber \\
 %&& =|D_x({\mathbf k})|\exp{(i\phi_{x}({\mathbf k}))},
 \label{Aycc}
 \\
\mathcal{A}_{x}^{vv}(\mathbf k)&=&\frac{-a\gamma ^2}{\gamma ^2 |f(\mathbf k)|^2+(\Delta_g/2+E_c)^2}
 \sin \frac{3ak_y}{2\sqrt{3}}\sin{\frac{ak_x}{2}}
\nonumber \\
 %&& =|D_x({\mathbf k})|\exp{(i\phi_{x}({\mathbf k}))},
 \label{Axvv}
 \\
\mathcal{A}_{y}^{vv}(\mathbf k)&=&\frac{a\gamma ^2}{\sqrt{3}(\gamma ^2 |f(\mathbf k)|^2+(\Delta_g/2+E_c)^2)}\nonumber\\
 &&\times \Big(\cos{ ak_x}-\cos{\frac{\sqrt{3}ak_y}{2}}\cos{\frac{ak_x}{2}}\Big)
\nonumber \\
 %&& =|D_x({\mathbf k})|\exp{(i\phi_{x}({\mathbf k}))},
 \label{Ayvv}
\end{eqnarray}
The general solution of TDSE (\ref{Sch}) can be expanded in the basis of the Houston functions as 
\begin{equation}
\Psi_{\bf q} ({\bf r},t)=\sum_{\alpha=c_1,c_2,v}\beta_{\alpha{\bf q}}(t) \Phi^\mathrm{(H)}_{\alpha {\bf q}}({\bf r},t),
\end{equation}
where $\beta_{\alpha{\bf q}}$ are the expansion coefficients, which  satisfy the following system of coupled differential equations
\begin{equation}
i\hbar\frac{\partial B_\mathbf q(t)}{\partial t}= H^\prime(\mathbf q,t){B_\mathbf q}(t)~.
\label{Schrodinger}
\end{equation}
Here the wave function (the vector of state) $B_q(t)$ and Hamiltonian in the interaction representation $ H^\prime(\mathbf q,t)$ are defined as 
\begin{eqnarray}
B_\mathbf q(t)&=&\begin{bmatrix}\beta_{c\mathbf q}(t)\\ \beta_{v\mathbf q}(t)\\ \end{bmatrix}~,\\ 
H^\prime(\mathbf q,t)&=&-e\mathbf F(t)\mathbfcal{\hat A}(\mathbf q,t)~,\\
\mathbfcal{\hat A}(\mathbf q,t)&=&\begin{bmatrix}0&\mathbfcal D^{cv}(\mathbf q,t)\\
\mathbfcal D^{vc}(\mathbf q,t)&0\\
\end{bmatrix}~.
\end{eqnarray}

%Note that fundamentally Eq.\ (\ref{eq:beta_1,2}) is a Schr\"odinger equation in the interaction representation in the adiabatic basis of the Houston functions, where the wave function is a two-component state vector $\left(\beta_{c\mathbf q},\beta_{v\mathbf q}\right)$. 

The non-Abelian Berry connection matrix elements, $\mathbfcal{A}^{cv}=(\mathcal{A}^{cv}_{x},\mathcal{A}^{cv}_{y})$, which are proportional to the interband dipole matrix elements, are given by the 
following expressions
\begin{eqnarray}
\mathcal{A}_{x}^{cv}(\mathbf k)&=&\mathcal N\Bigg(\frac{-a}{2|f(\mathbf k)|^2}\Bigg)\Bigg( \sin\frac{ak_x}{2}\sin\frac{a\sqrt{3}k_y}{2}
\nonumber\\
%\nonumber \\&&
&&+i \frac{\Delta_g}{2E_c}\Bigg(\cos \frac{a\sqrt{3}k_y}{2}\sin \frac{ak_x}{2}+\sin{ak_x}\Bigg)\Bigg)
\nonumber \\
 %&& =|D_x({\mathbf k})|\exp{(i\phi_{x}({\mathbf k}))},
 \label{Ax}
\\
\mathcal{A}_{y}^{cv}(\mathbf k)&=&\mathcal N\Bigg(\frac{a}{2\sqrt{3}|f(\mathbf k)|^2}\Bigg)\Bigg( -1-\cos\frac{a\sqrt{3}k_y}{2}\cos\frac{ak_x}{2}
\nonumber\\
%\nonumber \\&&
&&+2\cos ^2 \frac{ak_x}{2}-i \frac{3\Delta_g}{2E_c}\sin \frac{a\sqrt{3}k_y}{2}\cos \frac{ak_x}{2}\Bigg)
\nonumber \\
 %&&=|D_y({\mathbf k})|\exp{(i\phi_{y}({\mathbf k}))},
\label{Ay}
\end{eqnarray}
where
\begin{equation}
\mathcal N=\frac{\left|\gamma f(\mathbf k)\right|}{\sqrt{\frac{\Delta_g^2}{4}+\left|\gamma f(\mathbf k)\right|^2}}~.
\end{equation}

%In these terms, we obtain Schr\"odinger equation in the interaction representation in the adiabatic basis of the Houston functions \cite{Houston_PR_1940_Electron_Acceleration_in_Lattice} as
In the presence of a strong field, we solve the TDSE (\ref{Schrodinger}) with  initial conditions that the VB is initially fully
occupied and the CB is completely empty. From these solutions we can find the  
residual CB population and find the energy 
absorbed by graphene monolayer. The corresponding absorbance is defined as 
\begin{equation}
A= \frac{\int |\beta_{c\mathbf q}(t=\infty)|^2 E_{\mathrm c}(\mathbf q)d\mathrm{\mathbf q}}{2\pi^2c\epsilon_0\int_{-\infty}^{\infty}|\mathbf F|^2dt} ,
\label{A}
\end{equation}
%\textcolor{red}%
{where $c$ is speed of light in vacuum, and $\epsilon_0$ is the dielectric permittivity of the surrounding medium.}

Below we present the results for a gapped graphene monolayer for both linearly- and circularly-polarized pulses. 

\section{Results}
\subsection{Linearly polarized ultrafast pulse}
First, we consider a linearly-polarized optical pulse that consists of a single oscillation and is polarized along the $x$ axis, ${\mathbf {F}}(t)=(F_x(t),0)$. The waveform of this pulse is set as
\begin{eqnarray}
F_x(t)&=&F_0\left(1-2u^2\right)\exp{\left(-u^2\right)}
\label{Fx}
\end{eqnarray}
where $F_0$ is the amplitude of the pulse, and $\tau= 1$ fs is the characteristic time of the optical oscillation.
  
\begin{figure}
\begin{center}\includegraphics[width=0.47\textwidth]{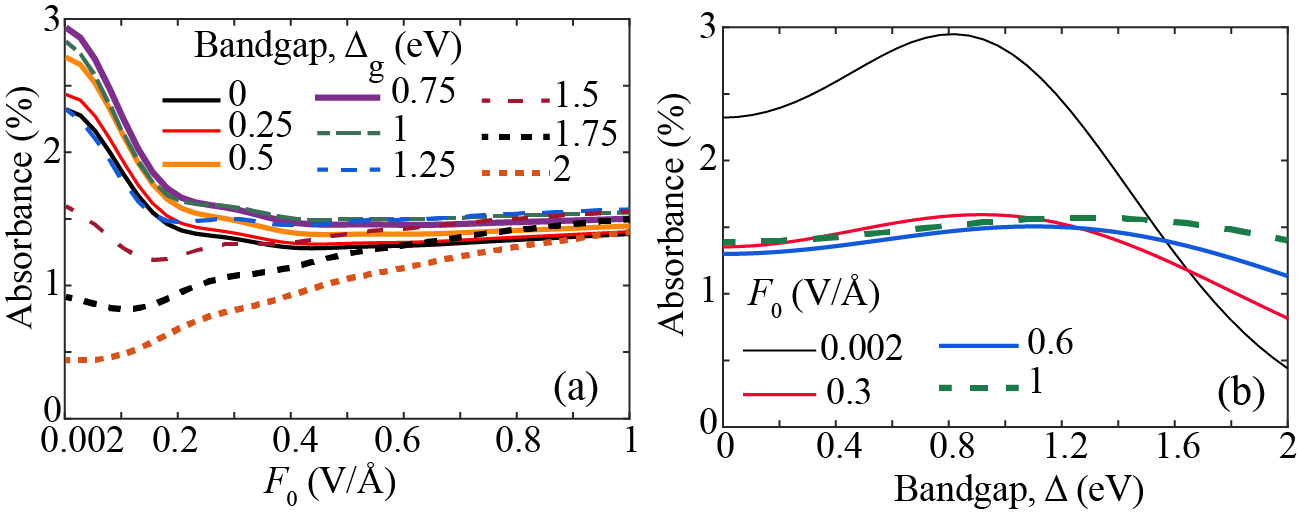}\end{center}
  \caption{(Color online) Absorbance  in gapped graphene  (a) as a function of the field amplitude for $\Delta_g=0,~0.25,~0.5,~ 0.75,~1,~1.25,~1.5,~1.75,~2~\mathrm{eV}$ and (b) as a function of bandgap for $F_0=0.002,~ 0.3,~0.6,~1~\mathrm{V/\AA}$. The applied optical pulse is linearly polarized along $x$ direction. 
 % \color{green}{The dimensionality seems to be wrong: Judging from Eq.\ (\ref{A}), $A$ has dimensionality $\mathrm{cm^2}$. However, in this Figure it is dimensionless. Please recheck twice.}
  }
  \label{fig:Absorption_1L_gap_vs_F0}
\end{figure}

%The amplitude of the applied pulse is several tenth of $\mathrm{V/\AA}$Universal optical absorption coefficient of graphene is .
The calculated absorbance as a 
function of the field amplitude is shown in Fig.\ \ref{fig:Absorption_1L_gap_vs_F0} for different values of the bandgap, $\Delta_g $. 
For the pristine graphene, $\Delta_g=0$, the absorbance takes the universal value of $\pi\alpha\approx 2.3 $ percent ($\alpha=\frac{1}{137}$ is the fine structure constant) for the field amplitudes as small as $0.002~\mathrm{V/\AA}$. 

\begin{figure}
\begin{center}\includegraphics[width=0.47\textwidth]{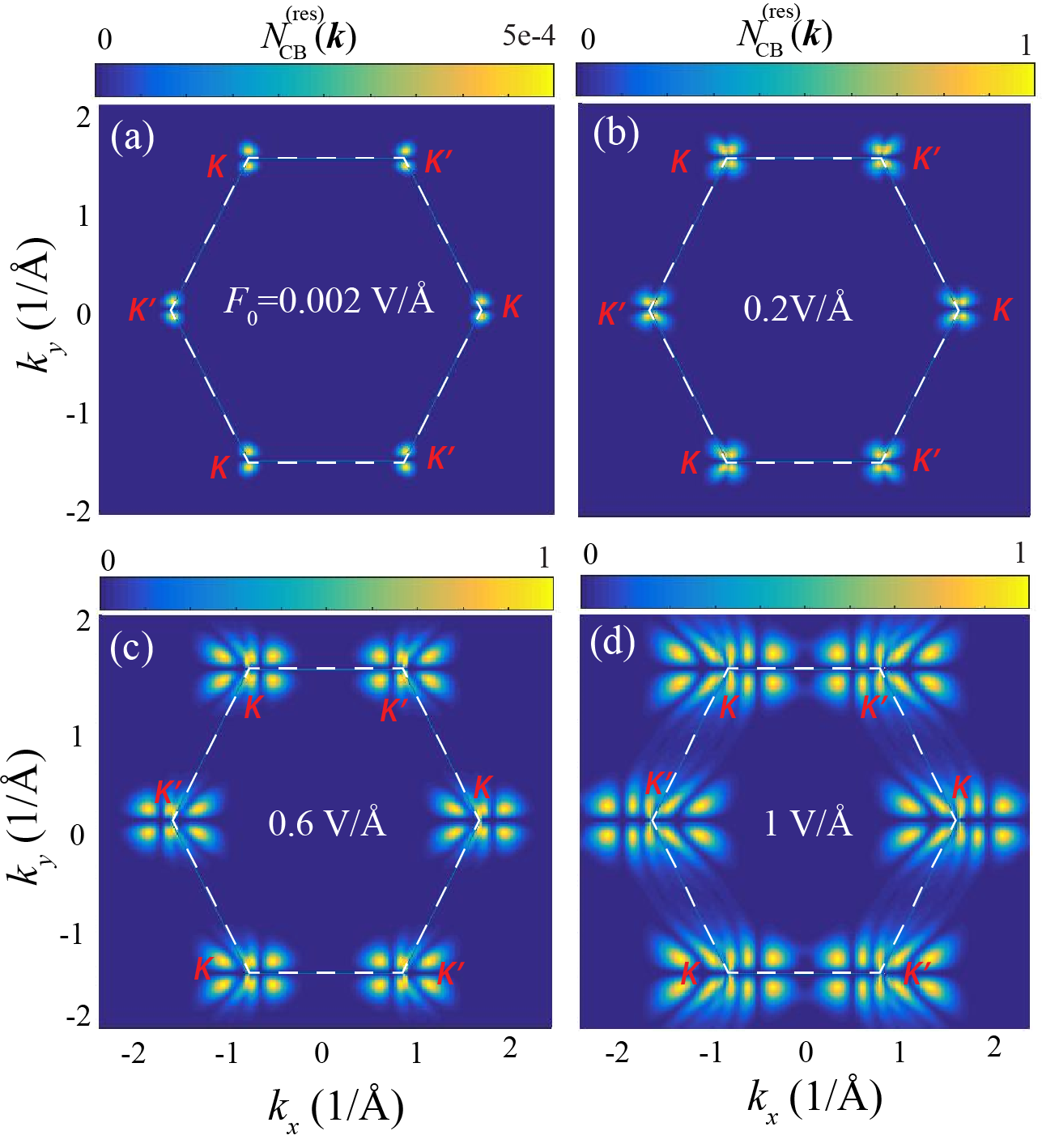}\end{center}
  \caption{(Color online)  Residual CB population of graphene with $\Delta_g=0$ induced by a linearly polarized pulse in $x$ direction. 
The field amplitude is (a) 0.002, (b) 0.2, (c) 0.6, and (D) 1 $\mathrm{V/\AA}$. The white dash line indicates the boundary of the first Brillouin zone.}
  \label{fig:L_gap_0_F0}
\end{figure}

With increasing the field amplitude, $F_0$, the absorbance  
decreases for small $\Delta_g $, $\Delta_g \lesssim 1.5$ eV, and increases for 
large $\Delta_g $, $\Delta_g \gtrsim 1.5$ eV. Finally, it reaches the saturated 
value of $\approx 1.5$  percent at $F_0\gtrsim 0.5$ V/\protect{\AA} -- see Fig.\ \ref{fig:Absorption_1L_gap_vs_F0}(a). 
Visible suppression of absorbance at small $\Delta_g$ with increasing the 
field amplitude can be understood by looking at the CB population distribution  for different field amplitudes 0.002, 0.2, 0.6, and 1 $\mathrm{V/\AA}$ -  see Fig.\ \ref{fig:L_gap_0_F0}, where the results are shown for pristine graphene.   
%%%%%%%%
%%%%%%%

The CB population distribution is determined by  the 
properties of the interband dipole matrix element (non-Abelian Berry 
connection). Namely, for pristine graphene, the $x$ component of the 
interband dipole matrix element 
has singularities at the Dirac points, $K$ and $K^\prime$, of type $q_y/(q_x^2+q_y^2)$, where $(q_x,q_y)$ is the wavevector defined relative to the corresponding 
Dirac point. Thus the interband coupling in this case is strongly localized 
at the Dirac points. As a result, if during the pulse an electron passes 
through the region that is very close to the Dirac point, then strongly localized 
interband coupling rotates the electron pseudospin by $180^0$. It means that
if before the passage the electron is in the VB then after the passage the electron is completely transferred to the CB and vice versa. The double passage of the Dirac point leaves the electron in the original state, i.e., in the VB.

Due to this property of the interband coupling, the CB population distribution has the following structure. For the field amplitude as small as 0.002 $\mathrm{V/\AA}$, near each Dirac point, there are two hot spots [see Fig.\ \ref{fig:L_gap_0_F0} (a)], 
i.e., regions with large CB population, one above and another below the 
Dirac point. They are separated by dark regions, which, as mentioned above, are due to singularity 
of the dipole matrix elements at the Dirac points. With increasing the field 
amplitude each bright spot transforms into a set of dark and bright fringes 
[see Fig.\ \ref{fig:L_gap_0_F0} (b)-(d)], which are due to interference effects.
%\textcolor{red}
{The first appearance of such an interference pattern occurs at a field amplitude of $F_0^{(1)}$ that can be calculated  as
\begin{equation}
F_0^{(1)} \approx 2\frac{\hbar\omega^2}{e v_F}~,
\label{F1}
\end{equation}
where $\omega$ is the characteristic carrier frequency of the pulse, and $v_F\sim\alpha c=e^2/\hbar$ is the Fermi velocity of electrons. For $\hbar\omega\approx 1.6$ eV, we obtain $F_0^{(1)} \approx 0.3$ V/\protect{\AA}. This is in a qualitative agreement with the calculation results illustrated in Fig.\ \ref{fig:L_gap_0_F0} (b)]. 
Note that the same field amplitude of   $\sim F_0^{(1)}$ determines the onset of the saturation of the absorbance -- cf.\ Fig.\ \ref{fig:Absorption_1L_gap_vs_F0}(a). Additionally, one can also evaluate the separation between the fringes in the reciprocal space, $\Delta k$, as
%%%%%%%%%%%%%%
\begin{equation}
\Delta k =\frac{2\omega}{v_F}~.
\label{Delta_k}
\end{equation}
Note that this separation does not depend on the pulse amplitude, which defines the number of fringes estimated as $\approx F_0/F_0^{(1)}$. Estimating from Eq.\ (\ref{Delta_k}), for  $\hbar\omega=1.6$ eV, we obtain $\Delta k\approx 0.2~\AA^{-1}$. This is in a good quantitative agreement with Fig.\ \ref{fig:L_gap_0_F0}(c), (d).
}

% As shown in Fig.\ \ref{fig:L_gap_0_F0} (a), for the field amplitude as low as 0.002 $\mathrm{V/\AA}$ there are two hot spots, regions with maximum population, upper and lower Dirac points where the Dirac points are indicated by $K$ and $K^\prime$ on the figure. Since dipole matrix element or interband Berry connection has singularity at Dirac points there are no electron population at these points. With increasing the field amplitude, the electron in the vicinity of the Dirac points travels a long distance in the reciprocal space and accumulates  a phase which is the reason of having interference pattern in the CB population distribution shown in Fig.\ \ref{fig:L_gap_0_F0} (b)-(D). Due to the existence of the interference pattern the number of hot spots increases with the amplitude of the applied pulse, as a resultthe absorption coefficient decrease with increasing the field amplitude -- see Fig.\ \ref{fig:Absorption_1L_gap_vs_F0} for $\Delta_g=0$

\begin{figure}
\begin{center}\includegraphics[width=0.47\textwidth]{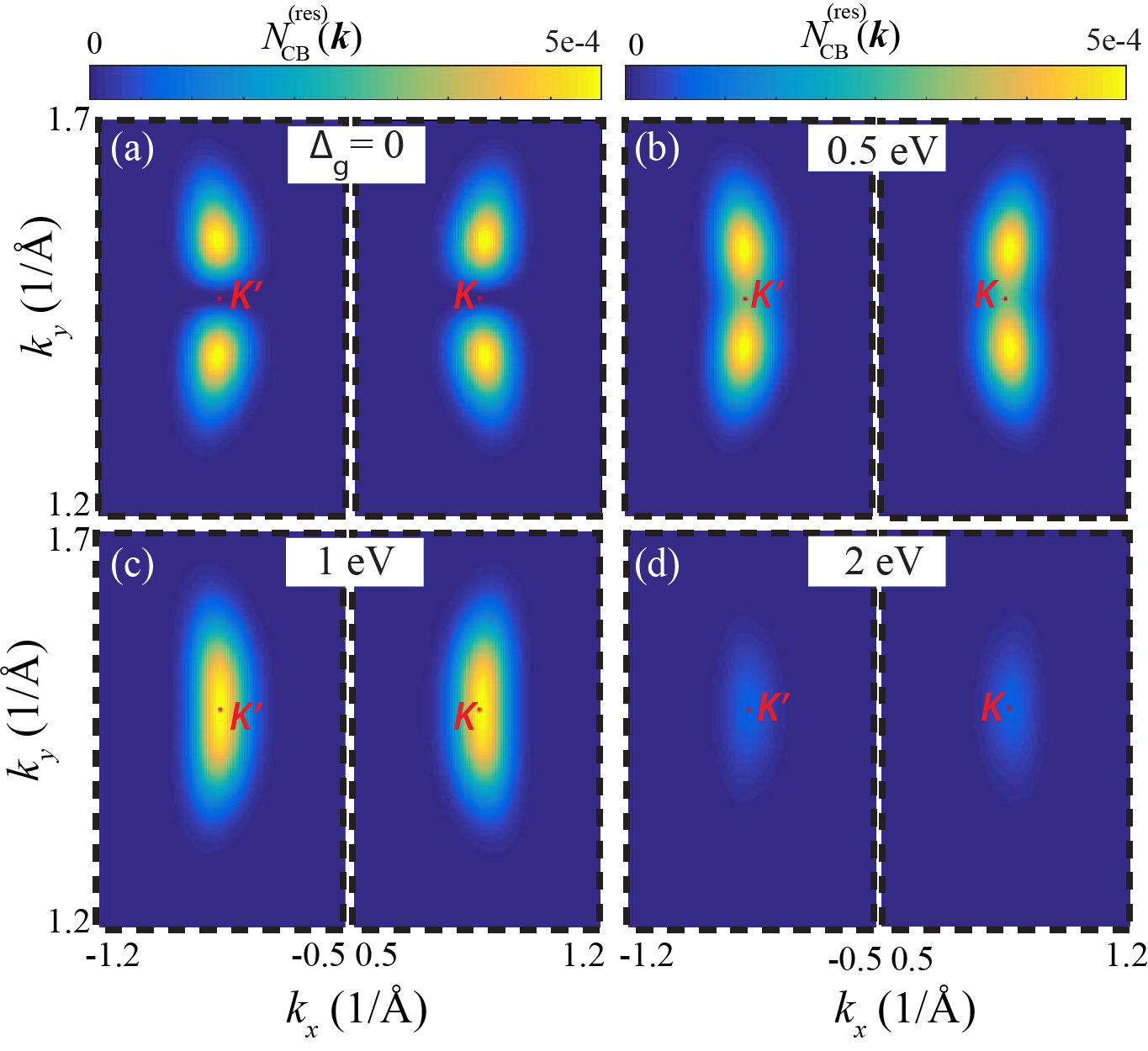}\end{center}
  \caption{(Color online) Residual CB population distribution for gapped graphene with the bandgap of $\Delta_g =0$, 0.5, 1, and 2 eV. 
The optical pulse is linearly polarized along $x$ direction. The amplitude 
of the pulse is  0.002 $\mathrm{V/\AA}$. The CB population distributions 
are shown near the $K$ and $K^\prime $ points. 
%\textcolor{green}{What parts of the Brillouin zone are shown and what is the dashed line at the boundaries?}
}
  \label{fig:L_F0_0p002_gap}
\end{figure}

As a function of the bandgap, $\Delta_g$, the absorbance  shows 
different types of behavior at small and large field amplitudes, $F_0$, -- see 
Fig.\  \ref{fig:Absorption_1L_gap_vs_F0}. At large $F_0$, $F_0 \gtrsim \Delta_g/a $, %\textcolor{red} 
{where $a$ is the lattice constant,} the absorbance  is almost independent of the bandgap, while at smaller $F_0$ the absorbance  has strong 
nonmonotonic a dependence on $\Delta_g $ -- see Fig.\  \ref{fig:Absorption_1L_gap_vs_F0}(b). The origin of such  a dependence  can be understood from the CB population distribution, which is shown in 
Fig.\ %\textcolor{red}
{\ref{fig:L_F0_0p002_gap}} for the field amplitude of $F_0 = 0.002$ V/\protect{\AA} and various bandgaps. 
For small $\Delta_g $, the CB population has two maxima above and below the $K$ and $K^{\prime }$ points. As mentioned above, in 
pristine graphene, i.e., at zero bandgap, these maxima are due to singularities of the interband dipole matrix element at the Dirac points.
At a finite bandgap, the interband coupling is regular and has a single maximum at 
each Dirac point with the maximum value that is inversely proportional to the 
bandgap. 
As a results, with increasing the bandgap, the CB population distribution
transforms from the two-maxima structure near each Dirac points into a single-maximum structure at the Dirac points, which occurs at $\Delta_g \approx 1$ eV.
In such a case, the absorbance  increases with $\Delta_g$ -- see Fig.\ \ref{fig:Absorption_1L_gap_vs_F0}(b). After that, when the 
bandgap increases further, the interband coupling at the Dirac point decreases, which suppresses both the CB population [see Fig.\ \ref{fig:L_F0_0p002_gap}(d)] and the absorbance. 

%%%%%%%%%%%%%%

%%%%%%%%%%%%%%
\begin{figure}
\begin{center}\includegraphics[width=0.47\textwidth]{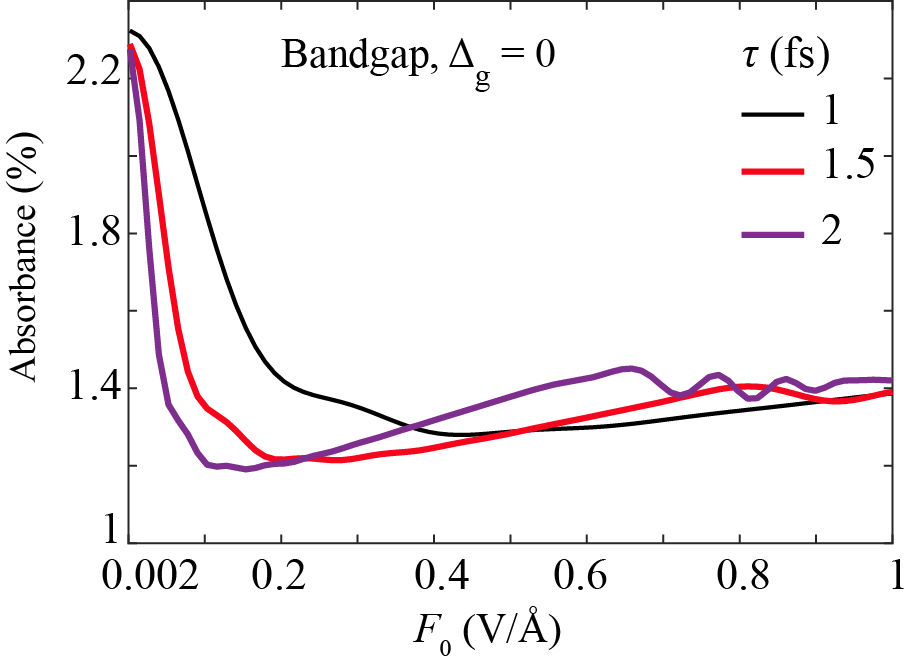}\end{center}
  \caption{(Color online)  Absorbance  of pristine graphene as a function of the field amplitude for different values of $\tau $, 
$\tau=$ 1, 1.5, 2 fs. 
}
  \label{fig:Absorption_CB_Linear_tau_1_1p5_2_fs_gap_0}
\end{figure}%
%%%%%%%%%%%%%

Another parameter of the pulse, which determines the absorbance  of the system, is the duration of the pulse, $\tau $ that determines its carrier frequency, $\omega\sim 1/\tau$. The 
dependence of the absorbance  on $\tau $ is shown in Fig.\ \ref{fig:Absorption_CB_Linear_tau_1_1p5_2_fs_gap_0} for pristine graphene with a
zero bandgap. The results are shown as a function of the field amplitude. 
We can see from the figure that the saturated value of the absorbance, which is $\approx 1.4$ percent and is achieved at large field amplitudes, does not depend on the duration of the pulse, $\tau $. 
At the same time, the field amplitude, at which the saturated value is 
achieved, depends on $\tau $. Namely, with increasing $\tau$ (or, decrease of $\omega\sim 1/\tau$, the saturated value is achieved at a smaller field amplitude $F_0$. 
%
%\textcolor{red}
{We can reasonably assume that the saturtion is also related to the formation of the interference fringes in the electron lattice-momentum distribution. In such a case, the field at which the saturation sets on is given by $F_0^{(1)}\propto 1/\tau^2$ -- see Eq.\ (\ref{F1}). This scaling is in a reasonable agreement with the numerical results of Fig.\ \ref{fig:Absorption_CB_Linear_tau_1_1p5_2_fs_gap_0}.
}

%This is because, as mentioned above, such field amplitude corresponds to the condition that the CB population distribution first shows the interference fringes. Such fringes occur when the electron trajectory in the reciprocal space due to the field of the pulse is long enough so the electrons during such transfer accumulate extra phase, which is enough to produce the interference pattern. The length of the trajectory is proportional to $\tau F_0$. Thus for large $\tau $ the corresponding field amplitude becomes small. Such abehavior is illustrated in Fig. \ref{fig:Absorption_CB_Linear_tau_1_1p5_2_fs_gap_0}.

% illustrates the effect of $\tau$ on the absorption coefficient curve of pristine graphene. $\tau$ identifies the mean frequency of the pulse and with increasing $\tau$ the mean frequency of the pulse is shifting to the lower frequencies while its bandwidth is fixed. As illustrated in this figure, independent from the value of $\tau$, the absorption coefficient is approximately equal to the universal value for low field amplitudes. For the higher field amplitudes the absorption coefficients reach to the saturation value of 1.4$ percent$ independently of the value of $\tau$. The speed of reduction in the absorption for a longer $\tau$ is explained with considering the fact that increment of $\tau$ leads to the larger electrons trajectory in the reciprocal space and acts similar to a higher field for a small value of $\tau$.

\subsection{Circularly polarized ultrafast pulse}

Here, we consider absorption of a single-oscillation circularly-polarized pulse. 
The profile of the pulse, $\mathbf{F}(t)=\{F_x(t),F_y(t)\}$, is defined by the following expressions  
\begin{eqnarray}
F_x(t)&=&F_0\left(1-2u^2\right)\exp{\left(-u^2\right)}
\label{Fx}
\\
F_y(t)&=&2F_0u\exp{\left(-u^2\right)}~,
\label{Fy}
\end{eqnarray}
where $F_0$ is the amplitude of the pulse and $u=t/\tau$.  

%For a given circular pulse, the electron trajectory is semicircle in the reciprocal space and CB states which are close to separatrix passes the region with strong interband peak is located in the vicinity of $K$ and $K^\prime$. Therefore the electron eigenstates accumulate some phases called a topological phase along the trajectory; this phase reveals the topology of energy band structure.

\begin{figure}
\begin{center}\includegraphics[width=0.47\textwidth]{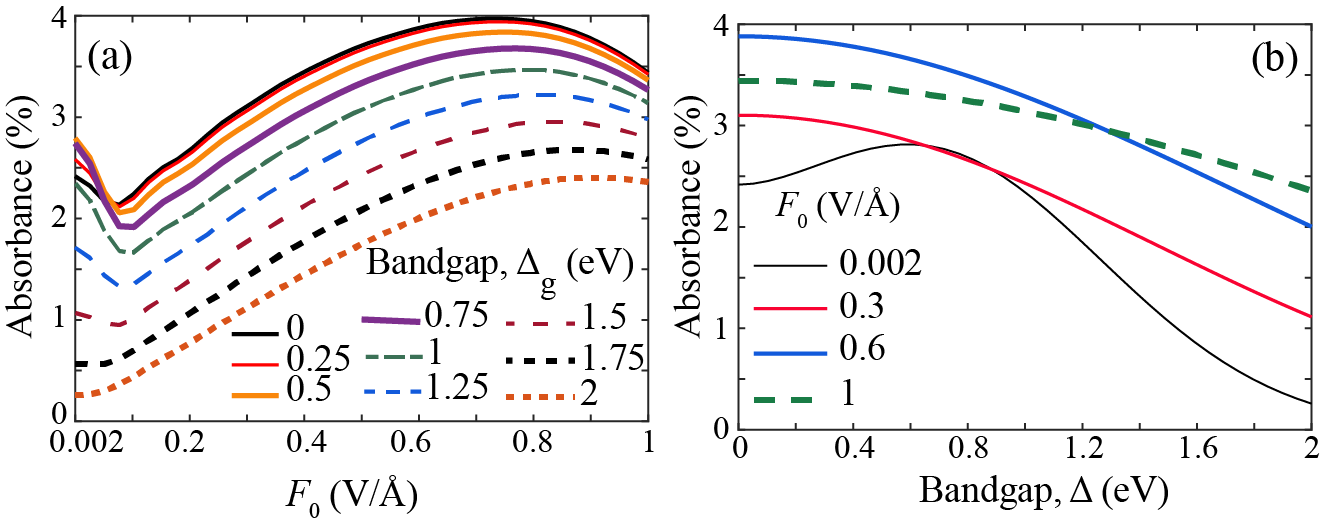}\end{center}
  \caption{(Color online) Absorbance  in gapped graphene with tunable bandgap (a) as a function of the field amplitude for $\Delta_g=0,~0.25,~0.5,~ 0.75,~1,~1.25,~1.5,~1.75,~2~\mathrm{eV}$ and (b) as a function of the bandgap for $F_0=0.002,~ 0.3,~0.6,~1~\mathrm{V/\AA}$. The  optical pulse is circularly polarized.}
  \label{fig:Absorption_CB_1C_gap_vs_F0}
\end{figure}

The absorbance  for a circularly polarized pulse is shown in 
Fig.\ \ref{fig:Absorption_CB_1C_gap_vs_F0}(a) as a function of the field amplitude, $F_0$, for different bandgaps. The absorbance  does not saturate 
at large values of $F_0$, in a sharp contrast to the case of the linearly 
polarized pulse. At all field amplitudes the absorbance  has strong 
dependence on the bandgap. For small $\Delta_g $, $\Delta_g \lesssim 1.5$ eV, 
the absorbance  first decreases with $F_0$, reaches its minimum value, 
and then increases. For large bandgaps, $\Delta_g \gtrsim 1.5$ eV, the 
absorbance  monotonically increases with $F_0$. 

\begin{figure}
\begin{center}\includegraphics[width=0.47\textwidth]{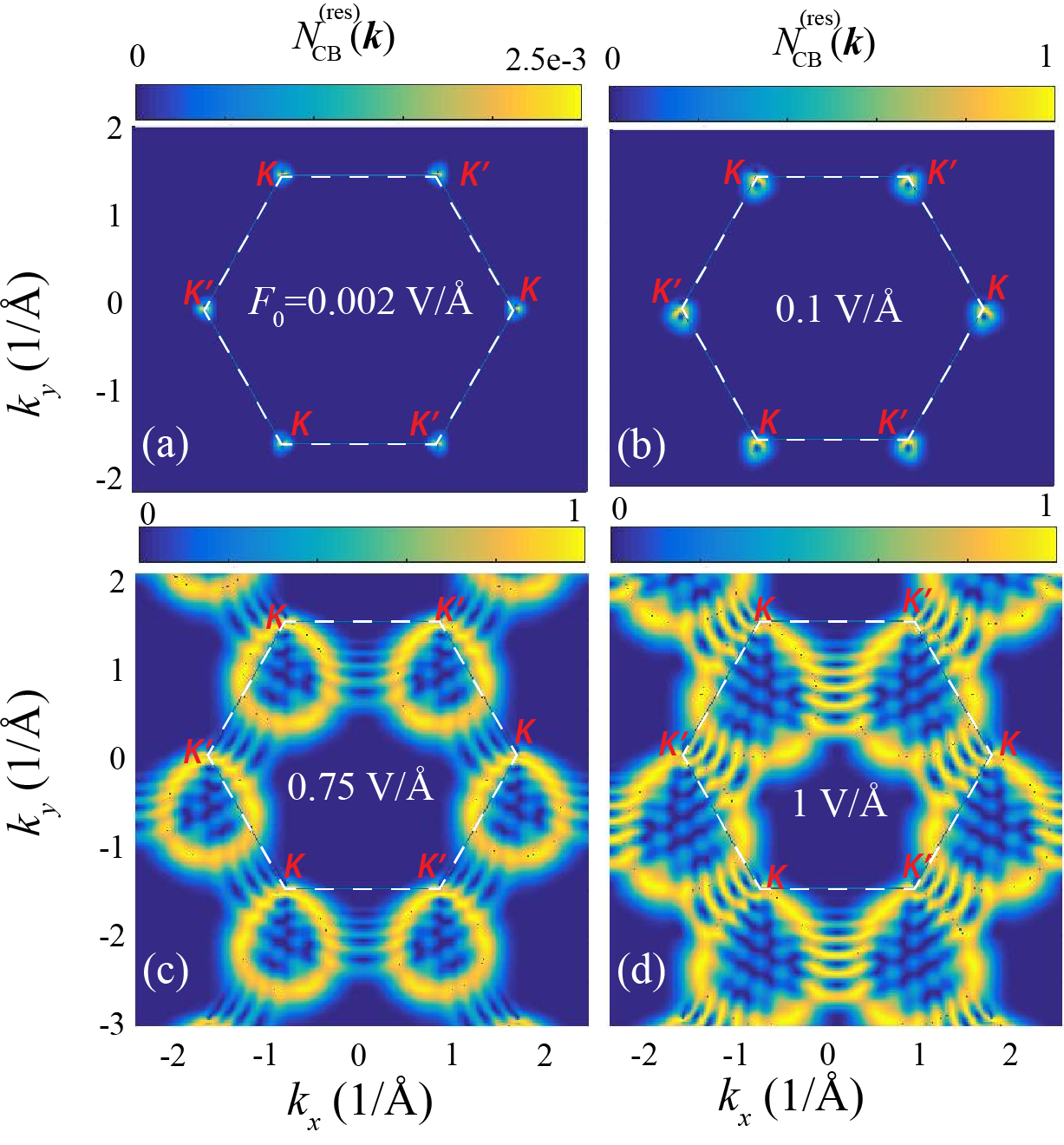}\end{center}
  \caption{(Color online)  Residual CB population of graphene with $\Delta_g=0$ induced by a circularly polarized pulse with different amplitudes 0.002, 0.1, 0.75, and 1 $\mathrm{V/\AA}$ for panels (a)-(d), correspondingly. The white dash line indicates the boundary of the first Brillouin zone. }
  \label{fig:1RC_gap_0_F0}
\end{figure}

The origin of nonmonotonic dependence of the absorbance  at small 
values of $\Delta_g$ can be understood from the CB population distribution shown in 
Fig. \ref{fig:1RC_gap_0_F0} for pristine graphene. At small field amplitudes, $F_0 \lesssim 0.1$ V/\AA, the CB population at each Dirac point, $K$ or $K^\prime$, has a single-peak structure localized at the corresponding Dirac point. Within this range of $F_0$, the absorbance  decreases with $F_0$. Then, at $F_0 \approx 0.1$ V/\protect{\AA}, 
a single peak structure of the CB population distribution transforms into an arc that is a caustic, i.e., an image of the separatrix whose size is proportional to $F_0$ -- see Refs.\  \onlinecite{Stockman_et_al_PhysRevB.98_2018_Rapid_Communication_Topological_Resonances, Stockman_et_al_PhysRevB.100.115431_2019_Gapped_Graphene}.
With this structure of the CB population, the absorbance  increases with $F_0$. Finally, the absorbance  reaches its maximum at $F_0 \approx 0.75$
V/\protect{\AA}. This is the value of $F_0$ at which the CB population distribution shows 
the first interference fringes -- see Fig. \ref{fig:1RC_gap_0_F0}(c). Such an interference 
pattern is clearly visible at large field amplitudes -- see  Fig. \ref{fig:1RC_gap_0_F0}(d). It is due to the Bloch trajectories crossing the $K,K^\prime$-valley boundaries, which is likely to limit the absorbance.

%%%%%%%%%
%%%%%%%%%%

%%%%%%%%%%%%
%%%%%%%%%%%%%%

\begin{figure}
\begin{center}\includegraphics[width=0.47\textwidth]{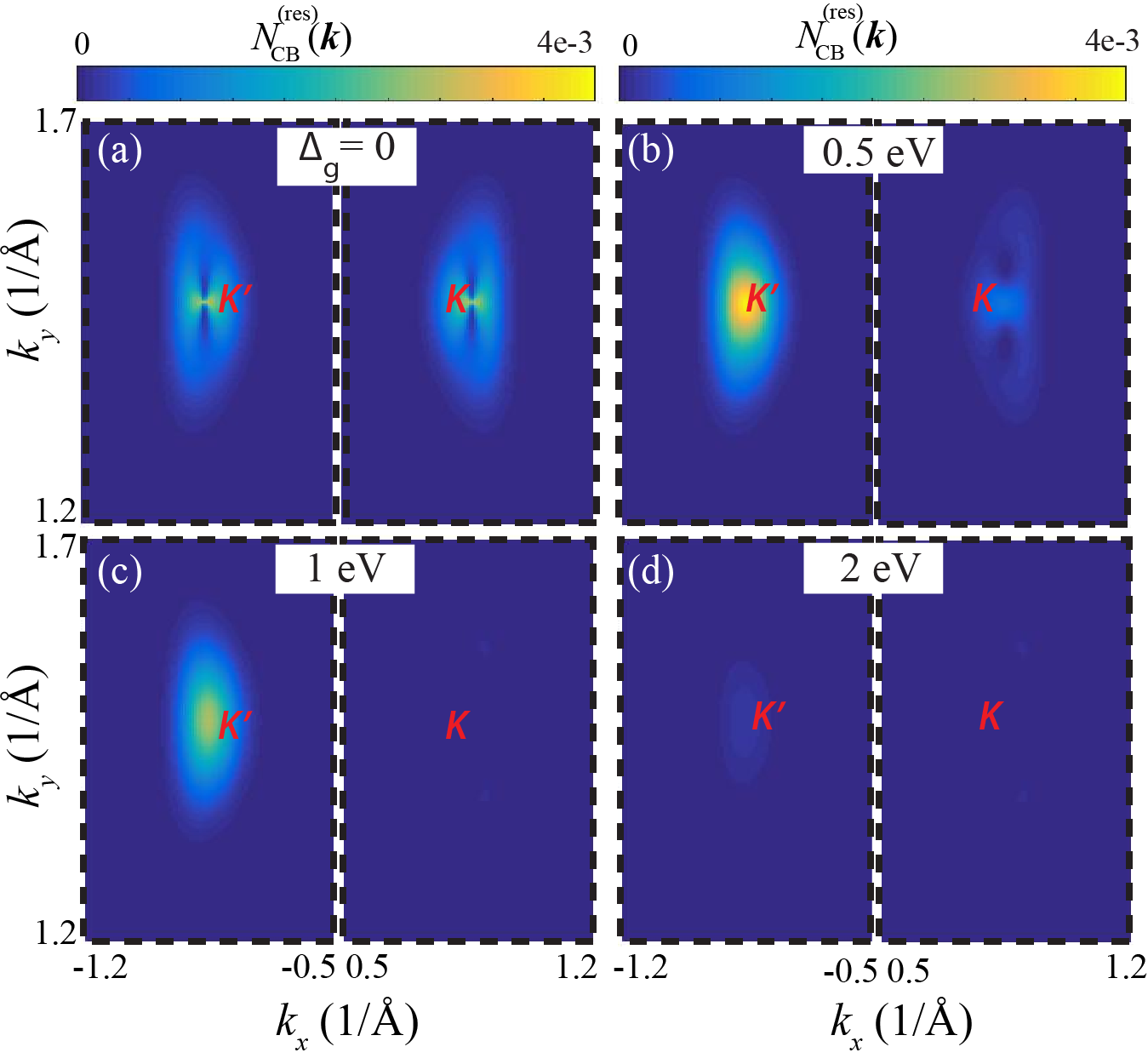}\end{center}
  \caption{(Color online) Residual CB population of gapped graphene induced by a circularly polarized pulse with the amplitude of 0.002 $\mathrm{V/\AA}$.  The bandgap 
is 0, 0.5, 1, and 2 eV, as indicated in the corresponding panels. The white dash line indicates the boundary of the first  Brillouin zone.}
  \label{fig:1RC_F0_0p002_gap}
\end{figure}

Similar to the case of a linearly polarized pulse, the absorbance  for a
circularly polarized pulse shows nonmonotonic dependence on the bandgap at small 
field amplitudes -- see Fig. \ref{fig:Absorption_CB_1C_gap_vs_F0}(b). Namely, the 
absorbance  first increases with $\Delta_g $ and then decreases. Such behavior is consistent with CB population distribution shown in Fig.\ \ref{fig:1RC_F0_0p002_gap} for the field amplitude of $F_0 = 0.002$ V/\AA. %The value of the CB population is determined by competition between two areas in the reciprocal space: (i) the area, $S_{F}$, enclosed by the electron trajectory in the field of the pulse and (ii) the area, $S_D$, within which the interband dipole matrix elements is large. For small $\Delta_g $, the dipole matrix element is strongly localized at the Dirac points and area $S_D$ is less then area $S_F$. In this case the CB population is large only within the area $S_D$. With increasing $\Delta_g$, the area $S_D$ increases, which results in larger CB population and correspondingly in increasing of the  absorbance. At bandgap of $\approx 0.8$ eV, both areas $S_F$ and $S_D$ are almost the same and this point corresponds to the maximum of the absorbance. With further increase of $\Delta_g $ the CB population decreases because within the area of $S_F$ the interband coupling becomes smaller. As a results the absorbance  decreases with $\Delta_g$. \textcolor{green}{These areas, $S_F$ and $S_D$ are not precisely defined or illustrated anywhere how they are positioned in the Brillouin zone. This may cause a confusion.}

%%%%%%%%%%%%%

%The distribution of the residual CB population of gapped graphene, $\Delta_g=0$, 0.5, 1, 2 eV is shown in in Fig.\ \ref{fig:1RC_F0_0p002_gap} for the applied circular pulse with amplitude 0.002 $\mathrm{V/\AA}$. This figure presents that for the weak fields CB population is around Dirac points and with increasing the bandgap valley polarization is well pronounced due to the recently predicted topological resonances and CB population decreases. 
%\begin{figure}
%\begin{center}\includegraphics[width=0.47\textwidth]{Absorption_CB_1C_F0_vs_gap.png}\end{center}
%  \caption{(Color online) Absorption as a function of the bandgap of gapped graphene. The results are shown for different field amplitudes of the circularly polarized intense laser pulse which has the maximum in the x direction ($F_0=$ 0.002, 0.3, 0.6, and 1 $\mathrm{V/\AA}$). }
%  \label{fig:Absorption_CB_1C_F0_vs_gap}
%\end{figure}%

\section{Conclusions}

The ultrafast absorption of optical pulses in gapped graphene is determined by 
specific properties of ultrafast electron dynamics, both intraband and interband, in the field of the pulse. Such dynamics strongly depends on polarization of the 
optical pulse, whether it is linear or circular.
 There is a fundamental difference between these two 
types of single-oscillation pulses. isIn fact,  the electron Bloch trajectory 
in the reciprocal space passes twice though the region near the $K,K^\prime$-points (where the interband coupling is large) for the linear polarization and only once for the circular polarization. As a result, the interference pattern with the dark and bright fringes is clearly visible in the CB population 
distribution for a linearly polarized pulse but no such interference is observed at small field amplitudes for a circularly polarized pulse in agreement with earlier results \cite{Stockman_et_al_PhysRevB.98_2018_Rapid_Communication_Topological_Resonances, Stockman_et_al_PhysRevB.100.115431_2019_Gapped_Graphene}. Due to this effect, the absorption of the linearly- and circularly-polarized pulses is different. 
Such a difference is well pronounced for relatively large field amplitudes, $F_0\gtrsim 0.1$ V/\AA. For small field amplitudes, the absorbance for both types of polarization behaves similarly. This is because for 
small field amplitudes the size of the electron displacement in the reciprocal space is less than or comparable to the size of the region with the large interband coupling. In this case, during the whole trajectory, both for linearly 
and circularly polarized pulses, there is a strong (or weak) interband coupling. Thus no interference pattern can be formed and no difference between the linear and circular polarizations is observed.

At large field amplitude, when the interference pattern is formed for linearly polarized pulses, the main differences between the circularly and linearly polarized 
pulse can be summarized as follows. 
While for linearly polarized pulse the absorbance as a function of the pulse 
amplitude is saturated at $\approx 1.4$ percent, for circularly polarized pulse the 
absorbance does not show any saturation. The absorbance of a circularly polarized pulse 
can reach the value of as much as 4  percent. 
As a function of the bandgap, the absorbance of a linearly polarized pulse has weak dependence on $\Delta_g$, while the absorbance of a circularly polarized pulse strongly depends on the bandgap.

\begin{acknowledgments}
Major funding was provided by Grant No. DE-FG02-11ER46789 from the Materials Sciences and Engineering Division of the Office of the Basic Energy Sciences, Office of Science, U.S. Department of Energy. Numerical simulations have been performed using support by Grant No. DE-FG02-01ER15213 from the Chemical Sciences, Biosciences and Geosciences Division, Office of Basic Energy Sciences, Office of Science, US Department of Energy. The work of V.A. was supported by NSF EFRI NewLAW Grant EFMA-17 41691.
\end{acknowledgments}

%\bibliography{../../BibTex/references}
%\bibliography{references}

\begin{thebibliography}{44}%
\makeatletter
\providecommand \@ifxundefined [1]{%
 \@ifx{#1\undefined}
}%
\providecommand \@ifnum [1]{%
 \ifnum #1\expandafter \@firstoftwo
 \else \expandafter \@secondoftwo
 \fi
}%
\providecommand \@ifx [1]{%
 \ifx #1\expandafter \@firstoftwo
 \else \expandafter \@secondoftwo
 \fi
}%
\providecommand \natexlab [1]{#1}%
\providecommand \enquote  [1]{``#1''}%
\providecommand \bibnamefont  [1]{#1}%
\providecommand \bibfnamefont [1]{#1}%
\providecommand \citenamefont [1]{#1}%
\providecommand \href@noop [0]{\@secondoftwo}%
\providecommand \href [0]{\begingroup \@sanitize@url \@href}%
\providecommand \@href[1]{\@@startlink{#1}\@@href}%
\providecommand \@@href[1]{\endgroup#1\@@endlink}%
\providecommand \@sanitize@url [0]{\catcode `\\12\catcode `\$12\catcode
  `\&12\catcode `\#12\catcode `\^12\catcode `\_12\catcode `\%12\relax}%
\providecommand \@@startlink[1]{}%
\providecommand \@@endlink[0]{}%
\providecommand \url  [0]{\begingroup\@sanitize@url \@url }%
\providecommand \@url [1]{\endgroup\@href {#1}{\urlprefix }}%
\providecommand \urlprefix  [0]{URL }%
\providecommand \Eprint [0]{\href }%
\providecommand \doibase [0]{http://dx.doi.org/}%
\providecommand \selectlanguage [0]{\@gobble}%
\providecommand \bibinfo  [0]{\@secondoftwo}%
\providecommand \bibfield  [0]{\@secondoftwo}%
\providecommand \translation [1]{[#1]}%
\providecommand \BibitemOpen [0]{}%
\providecommand \bibitemStop [0]{}%
\providecommand \bibitemNoStop [0]{.\EOS\space}%
\providecommand \EOS [0]{\spacefactor3000\relax}%
\providecommand \BibitemShut  [1]{\csname bibitem#1\endcsname}%
\let\auto@bib@innerbib\@empty
%</preamble>
\bibitem [{\citenamefont {Schiffrin}\ \emph {et~al.}(2012)\citenamefont
  {Schiffrin}, \citenamefont {Paasch-Colberg}, \citenamefont {Karpowicz},
  \citenamefont {Apalkov}, \citenamefont {Gerster}, \citenamefont {Muhlbrandt},
  \citenamefont {Korbman}, \citenamefont {Reichert}, \citenamefont {Schultze},
  \citenamefont {Holzner}, \citenamefont {Barth}, \citenamefont {Kienberger},
  \citenamefont {Ernstorfer}, \citenamefont {Yakovlev}, \citenamefont
  {Stockman},\ and\ \citenamefont
  {Krausz}}]{Schiffrin_at_al_Nature_2012_Current_in_Dielectric}%
  \BibitemOpen
  \bibfield  {author} {\bibinfo {author} {\bibfnamefont {A.}~\bibnamefont
  {Schiffrin}}, \bibinfo {author} {\bibfnamefont {T.}~\bibnamefont
  {Paasch-Colberg}}, \bibinfo {author} {\bibfnamefont {N.}~\bibnamefont
  {Karpowicz}}, \bibinfo {author} {\bibfnamefont {V.}~\bibnamefont {Apalkov}},
  \bibinfo {author} {\bibfnamefont {D.}~\bibnamefont {Gerster}}, \bibinfo
  {author} {\bibfnamefont {S.}~\bibnamefont {Muhlbrandt}}, \bibinfo {author}
  {\bibfnamefont {M.}~\bibnamefont {Korbman}}, \bibinfo {author} {\bibfnamefont
  {J.}~\bibnamefont {Reichert}}, \bibinfo {author} {\bibfnamefont
  {M.}~\bibnamefont {Schultze}}, \bibinfo {author} {\bibfnamefont
  {S.}~\bibnamefont {Holzner}}, \bibinfo {author} {\bibfnamefont {J.~V.}\
  \bibnamefont {Barth}}, \bibinfo {author} {\bibfnamefont {R.}~\bibnamefont
  {Kienberger}}, \bibinfo {author} {\bibfnamefont {R.}~\bibnamefont
  {Ernstorfer}}, \bibinfo {author} {\bibfnamefont {V.~S.}\ \bibnamefont
  {Yakovlev}}, \bibinfo {author} {\bibfnamefont {M.~I.}\ \bibnamefont
  {Stockman}}, \ and\ \bibinfo {author} {\bibfnamefont {F.}~\bibnamefont
  {Krausz}},\ }\bibfield  {title} {\enquote {\bibinfo {title}
  {Optical-field-induced current in dielectrics},}\ }\href@noop {} {\bibfield
  {journal} {\bibinfo  {journal} {Nature}\ }\textbf {\bibinfo {volume} {493}},\
  \bibinfo {pages} {70--74} (\bibinfo {year} {2012})}\BibitemShut {NoStop}%
\bibitem [{\citenamefont {Apalkov}\ and\ \citenamefont
  {Stockman}(2012)}]{Apalkov_Stockman_PRB_2012_Strong_Field_Reflection}%
  \BibitemOpen
  \bibfield  {author} {\bibinfo {author} {\bibfnamefont {V.}~\bibnamefont
  {Apalkov}}\ and\ \bibinfo {author} {\bibfnamefont {M.~I.}\ \bibnamefont
  {Stockman}},\ }\bibfield  {title} {\enquote {\bibinfo {title} {Theory of
  dielectric nanofilms in strong ultrafast optical fields},}\ }\href@noop {}
  {\bibfield  {journal} {\bibinfo  {journal} {Phys. Rev. B}\ }\textbf {\bibinfo
  {volume} {86}},\ \bibinfo {pages} {165118--1--13} (\bibinfo {year}
  {2012})}\BibitemShut {NoStop}%
\bibitem [{\citenamefont {Higuchi}\ \emph
  {et~al.}(2017{\natexlab{a}})\citenamefont {Higuchi}, \citenamefont {Heide},
  \citenamefont {Ullmann}, \citenamefont {Weber},\ and\ \citenamefont
  {Hommelhoff}}]{Higuchi_Hommelhoff_et_al_Nature_2017_Currents_in_Graphene}%
  \BibitemOpen
  \bibfield  {author} {\bibinfo {author} {\bibfnamefont {T.}~\bibnamefont
  {Higuchi}}, \bibinfo {author} {\bibfnamefont {C.}~\bibnamefont {Heide}},
  \bibinfo {author} {\bibfnamefont {K.}~\bibnamefont {Ullmann}}, \bibinfo
  {author} {\bibfnamefont {H.~B.}\ \bibnamefont {Weber}}, \ and\ \bibinfo
  {author} {\bibfnamefont {P.}~\bibnamefont {Hommelhoff}},\ }\bibfield  {title}
  {\enquote {\bibinfo {title} {Light-field-driven currents in graphene},}\
  }\href@noop {} {\bibfield  {journal} {\bibinfo  {journal} {Nature}\ }\textbf
  {\bibinfo {volume} {550}},\ \bibinfo {pages} {224--228} (\bibinfo {year}
  {2017}{\natexlab{a}})}\BibitemShut {NoStop}%
\bibitem [{\citenamefont {Gruber}\ \emph {et~al.}(2016)\citenamefont {Gruber},
  \citenamefont {Wilhelm}, \citenamefont {Pétuya}, \citenamefont {Smejkal},
  \citenamefont {Kozubek}, \citenamefont {Hierzenberger}, \citenamefont
  {Bayer}, \citenamefont {Aldazabal}, \citenamefont {Kazansky}, \citenamefont
  {Libisch}, \citenamefont {Krasheninnikov}, \citenamefont {Schleberger},
  \citenamefont {Facsko}, \citenamefont {Borisov}, \citenamefont {Arnau},\ and\
  \citenamefont
  {Aumayr}}]{Gruber_et_al_ncomms13948_2016_Ultrafast_pulses_graphene}%
  \BibitemOpen
  \bibfield  {author} {\bibinfo {author} {\bibfnamefont {Elisabeth}\
  \bibnamefont {Gruber}}, \bibinfo {author} {\bibfnamefont {Richard~A.}\
  \bibnamefont {Wilhelm}}, \bibinfo {author} {\bibfnamefont {Rémi}\
  \bibnamefont {Pétuya}}, \bibinfo {author} {\bibfnamefont {Valerie}\
  \bibnamefont {Smejkal}}, \bibinfo {author} {\bibfnamefont {Roland}\
  \bibnamefont {Kozubek}}, \bibinfo {author} {\bibfnamefont {Anke}\
  \bibnamefont {Hierzenberger}}, \bibinfo {author} {\bibfnamefont
  {Bernhard~C.}\ \bibnamefont {Bayer}}, \bibinfo {author} {\bibfnamefont
  {Iñigo}\ \bibnamefont {Aldazabal}}, \bibinfo {author} {\bibfnamefont
  {Andrey~K.}\ \bibnamefont {Kazansky}}, \bibinfo {author} {\bibfnamefont
  {Florian}\ \bibnamefont {Libisch}}, \bibinfo {author} {\bibfnamefont
  {Arkady~V.}\ \bibnamefont {Krasheninnikov}}, \bibinfo {author} {\bibfnamefont
  {Marika}\ \bibnamefont {Schleberger}}, \bibinfo {author} {\bibfnamefont
  {Stefan}\ \bibnamefont {Facsko}}, \bibinfo {author} {\bibfnamefont
  {Andrei~G.}\ \bibnamefont {Borisov}}, \bibinfo {author} {\bibfnamefont
  {Andrés}\ \bibnamefont {Arnau}}, \ and\ \bibinfo {author} {\bibfnamefont
  {Friedrich}\ \bibnamefont {Aumayr}},\ }\bibfield  {title} {\enquote {\bibinfo
  {title} {Ultrafast electronic response of graphene to a strong and localized
  electric field},}\ }\href {\doibase 10.1038/ncomms13948} {\bibfield
  {journal} {\bibinfo  {journal} {Nat. Commun.}\ }\textbf {\bibinfo {volume}
  {7}},\ \bibinfo {pages} {13948} (\bibinfo {year} {2016})}\BibitemShut
  {NoStop}%
\bibitem [{\citenamefont {Motlagh}\ \emph {et~al.}(2017)\citenamefont
  {Motlagh}, \citenamefont {Apalkov},\ and\ \citenamefont
  {Stockman}}]{Stockman_et_al_PhysRevB.95_2017_Crystalline_TI}%
  \BibitemOpen
  \bibfield  {author} {\bibinfo {author} {\bibfnamefont {S.~A.~Oliaei}\
  \bibnamefont {Motlagh}}, \bibinfo {author} {\bibfnamefont {V.}~\bibnamefont
  {Apalkov}}, \ and\ \bibinfo {author} {\bibfnamefont {M.~I.}\ \bibnamefont
  {Stockman}},\ }\bibfield  {title} {\enquote {\bibinfo {title} {Interaction of
  crystalline topological insulator with an ultrashort laser pulse},}\
  }\href@noop {} {\bibfield  {journal} {\bibinfo  {journal} {Phys. Rev. B}\
  }\textbf {\bibinfo {volume} {95}},\ \bibinfo {pages} {085438--1--8} (\bibinfo
  {year} {2017})}\BibitemShut {NoStop}%
\bibitem [{\citenamefont {Motlagh}\ \emph
  {et~al.}(2018{\natexlab{a}})\citenamefont {Motlagh}, \citenamefont {Wu},
  \citenamefont {Apalkov},\ and\ \citenamefont
  {Stockman}}]{Stockman_et_al_PhysRevB.98_2018_3D_TI}%
  \BibitemOpen
  \bibfield  {author} {\bibinfo {author} {\bibfnamefont {S.~A.~O.}\
  \bibnamefont {Motlagh}}, \bibinfo {author} {\bibfnamefont {J.~S.}\
  \bibnamefont {Wu}}, \bibinfo {author} {\bibfnamefont {V.}~\bibnamefont
  {Apalkov}}, \ and\ \bibinfo {author} {\bibfnamefont {M.~I.}\ \bibnamefont
  {Stockman}},\ }\bibfield  {title} {\enquote {\bibinfo {title} {Fundamentally
  fastest optical processes at the surface of a topological insulator},}\
  }\href {\doibase 10.1103/PhysRevB.98.125410} {\bibfield  {journal} {\bibinfo
  {journal} {Phys. Rev. B}\ }\textbf {\bibinfo {volume} {98}},\ \bibinfo
  {pages} {125410--1--11} (\bibinfo {year} {2018}{\natexlab{a}})}\BibitemShut
  {NoStop}%
\bibitem [{\citenamefont {Heide}\ \emph {et~al.}(2018)\citenamefont {Heide},
  \citenamefont {Higuchi}, \citenamefont {Weber},\ and\ \citenamefont
  {Hommelhoff}}]{Hommelhoff_et_al_PhysRevLett.121_2018_Coherent}%
  \BibitemOpen
  \bibfield  {author} {\bibinfo {author} {\bibfnamefont {C.}~\bibnamefont
  {Heide}}, \bibinfo {author} {\bibfnamefont {T.}~\bibnamefont {Higuchi}},
  \bibinfo {author} {\bibfnamefont {H.~B.}\ \bibnamefont {Weber}}, \ and\
  \bibinfo {author} {\bibfnamefont {P.}~\bibnamefont {Hommelhoff}},\ }\bibfield
   {title} {\enquote {\bibinfo {title} {Coherent electron trajectory control in
  graphene},}\ }\href@noop {} {\bibfield  {journal} {\bibinfo  {journal} {Phys.
  Rev. Lett.}\ }\textbf {\bibinfo {volume} {121}},\ \bibinfo {pages}
  {207401--1--5} (\bibinfo {year} {2018})}\BibitemShut {NoStop}%
\bibitem [{\citenamefont {Heide}\ \emph {et~al.}(2019)\citenamefont {Heide},
  \citenamefont {Boolakee}, \citenamefont {Higuchi}, \citenamefont {Weber},\
  and\ \citenamefont
  {Hommelhoff}}]{Hommelhoff_et_al_1903.07558_2019_laser_pulses_graphene}%
  \BibitemOpen
  \bibfield  {author} {\bibinfo {author} {\bibfnamefont {Christian}\
  \bibnamefont {Heide}}, \bibinfo {author} {\bibfnamefont {Tobias}\
  \bibnamefont {Boolakee}}, \bibinfo {author} {\bibfnamefont {Takuya}\
  \bibnamefont {Higuchi}}, \bibinfo {author} {\bibfnamefont {Heiko~B}\
  \bibnamefont {Weber}}, \ and\ \bibinfo {author} {\bibfnamefont {Peter}\
  \bibnamefont {Hommelhoff}},\ }\bibfield  {title} {\enquote {\bibinfo {title}
  {Interaction of carrier envelope phase-stable laser pulses with graphene: the
  transition from the weak-field to the strong-field regime},}\ }\href
  {\doibase 10.1088/1367-2630/ab13ce} {\bibfield  {journal} {\bibinfo
  {journal} {New J. Phys.}\ }\textbf {\bibinfo {volume} {21}},\ \bibinfo
  {pages} {045003} (\bibinfo {year} {2019})}\BibitemShut {NoStop}%
\bibitem [{\citenamefont {Sun}\ \emph {et~al.}(2012)\citenamefont {Sun},
  \citenamefont {Aivazian}, \citenamefont {Jones}, \citenamefont {Ross},
  \citenamefont {Yao}, \citenamefont {Cobden},\ and\ \citenamefont
  {Xu}}]{sun_et_al_nnano.2011.243_2012_Ultrafast_pulses_graphene}%
  \BibitemOpen
  \bibfield  {author} {\bibinfo {author} {\bibfnamefont {Dong}\ \bibnamefont
  {Sun}}, \bibinfo {author} {\bibfnamefont {Grant}\ \bibnamefont {Aivazian}},
  \bibinfo {author} {\bibfnamefont {Aaron~M.}\ \bibnamefont {Jones}}, \bibinfo
  {author} {\bibfnamefont {Jason~S.}\ \bibnamefont {Ross}}, \bibinfo {author}
  {\bibfnamefont {Wang}\ \bibnamefont {Yao}}, \bibinfo {author} {\bibfnamefont
  {David}\ \bibnamefont {Cobden}}, \ and\ \bibinfo {author} {\bibfnamefont
  {Xiaodong}\ \bibnamefont {Xu}},\ }\bibfield  {title} {\enquote {\bibinfo
  {title} {Ultrafast hot-carrier-dominated photocurrent in graphene},}\ }\href
  {\doibase 10.1038/nnano.2011.243
  https://www.nature.com/articles/nnano.2011.243#supplementary-information}
  {\bibfield  {journal} {\bibinfo  {journal} {Nat. Nanotechnol.}\ }\textbf
  {\bibinfo {volume} {7}},\ \bibinfo {pages} {114} (\bibinfo {year}
  {2012})}\BibitemShut {NoStop}%
\bibitem [{\citenamefont {Mashiko}\ \emph {et~al.}(2018)\citenamefont
  {Mashiko}, \citenamefont {Chisuga}, \citenamefont {Katayama}, \citenamefont
  {Oguri}, \citenamefont {Masuda}, \citenamefont {Takeda},\ and\ \citenamefont
  {Gotoh}}]{Mashiko_et_al_Nature_Communications_2018_ultrafast_pulse_solid}%
  \BibitemOpen
  \bibfield  {author} {\bibinfo {author} {\bibfnamefont {Hiroki}\ \bibnamefont
  {Mashiko}}, \bibinfo {author} {\bibfnamefont {Yuta}\ \bibnamefont {Chisuga}},
  \bibinfo {author} {\bibfnamefont {Ikufumi}\ \bibnamefont {Katayama}},
  \bibinfo {author} {\bibfnamefont {Katsuya}\ \bibnamefont {Oguri}}, \bibinfo
  {author} {\bibfnamefont {Hiroyuki}\ \bibnamefont {Masuda}}, \bibinfo {author}
  {\bibfnamefont {Jun}\ \bibnamefont {Takeda}}, \ and\ \bibinfo {author}
  {\bibfnamefont {Hideki}\ \bibnamefont {Gotoh}},\ }\bibfield  {title}
  {\enquote {\bibinfo {title} {Multi-petahertz electron interference in
  cr:al2o3 solid-state material},}\ }\href {\doibase
  10.1038/s41467-018-03885-7} {\bibfield  {journal} {\bibinfo  {journal} {Nat.
  Commun.}\ }\textbf {\bibinfo {volume} {9}},\ \bibinfo {pages} {1468}
  (\bibinfo {year} {2018})}\BibitemShut {NoStop}%
\bibitem [{\citenamefont {Shin}\ \emph {et~al.}(2018)\citenamefont {Shin},
  \citenamefont {Nguyen}, \citenamefont {Lim},\ and\ \citenamefont
  {Son}}]{Shin_et_al_IOP_Publishing_2018_ultrafast_pulse_solid}%
  \BibitemOpen
  \bibfield  {author} {\bibinfo {author} {\bibfnamefont {Hee~Jun}\ \bibnamefont
  {Shin}}, \bibinfo {author} {\bibfnamefont {Van~Luan}\ \bibnamefont {Nguyen}},
  \bibinfo {author} {\bibfnamefont {Seong~Chu}\ \bibnamefont {Lim}}, \ and\
  \bibinfo {author} {\bibfnamefont {Joo-Hiuk}\ \bibnamefont {Son}},\ }\bibfield
   {title} {\enquote {\bibinfo {title} {Ultrafast nonlinear travel of hot
  carriers driven by high-field terahertz pulse},}\ }\href {\doibase
  10.1088/1361-6455/aac59a} {\bibfield  {journal} {\bibinfo  {journal} {J.
  Phys. B: At. Mol. Opt. Phys.}\ }\textbf {\bibinfo {volume} {51}},\ \bibinfo
  {pages} {144003} (\bibinfo {year} {2018})}\BibitemShut {NoStop}%
\bibitem [{\citenamefont {Higuchi}\ \emph
  {et~al.}(2017{\natexlab{b}})\citenamefont {Higuchi}, \citenamefont {Heide},
  \citenamefont {Ullmann}, \citenamefont {Weber},\ and\ \citenamefont
  {Hommelhoff}}]{Higuchi_et_al_Nature_2017}%
  \BibitemOpen
  \bibfield  {author} {\bibinfo {author} {\bibfnamefont {Takuya}\ \bibnamefont
  {Higuchi}}, \bibinfo {author} {\bibfnamefont {Christian}\ \bibnamefont
  {Heide}}, \bibinfo {author} {\bibfnamefont {Konrad}\ \bibnamefont {Ullmann}},
  \bibinfo {author} {\bibfnamefont {Heiko~B.}\ \bibnamefont {Weber}}, \ and\
  \bibinfo {author} {\bibfnamefont {Peter}\ \bibnamefont {Hommelhoff}},\
  }\bibfield  {title} {\enquote {\bibinfo {title} {Light-field-driven currents
  in graphene},}\ }\href@noop {} {\bibfield  {journal} {\bibinfo  {journal}
  {Nature}\ }\textbf {\bibinfo {volume} {550}},\ \bibinfo {pages} {224--228}
  (\bibinfo {year} {2017}{\natexlab{b}})}\BibitemShut {NoStop}%
\bibitem [{\citenamefont {Trushin}\ \emph {et~al.}(2015)\citenamefont
  {Trushin}, \citenamefont {Grupp}, \citenamefont {Soavi}, \citenamefont
  {Budweg}, \citenamefont {Fazio}, \citenamefont {Sassi}, \citenamefont
  {Lombardo}, \citenamefont {Ferrari}, \citenamefont {Belzig}, \citenamefont
  {Leitenstorfer},\ and\ \citenamefont
  {Brida}}]{Leitenstorfer_et_al_PhysRevB.92_2015_Ultrafast_Pseudospin_Dynamics_in_Graphene}%
  \BibitemOpen
  \bibfield  {author} {\bibinfo {author} {\bibfnamefont {M.}~\bibnamefont
  {Trushin}}, \bibinfo {author} {\bibfnamefont {A.}~\bibnamefont {Grupp}},
  \bibinfo {author} {\bibfnamefont {G.}~\bibnamefont {Soavi}}, \bibinfo
  {author} {\bibfnamefont {A.}~\bibnamefont {Budweg}}, \bibinfo {author}
  {\bibfnamefont {D.~De}\ \bibnamefont {Fazio}}, \bibinfo {author}
  {\bibfnamefont {U.}~\bibnamefont {Sassi}}, \bibinfo {author} {\bibfnamefont
  {A.}~\bibnamefont {Lombardo}}, \bibinfo {author} {\bibfnamefont {A.~C.}\
  \bibnamefont {Ferrari}}, \bibinfo {author} {\bibfnamefont {W.}~\bibnamefont
  {Belzig}}, \bibinfo {author} {\bibfnamefont {A.}~\bibnamefont
  {Leitenstorfer}}, \ and\ \bibinfo {author} {\bibfnamefont {D.}~\bibnamefont
  {Brida}},\ }\bibfield  {title} {\enquote {\bibinfo {title} {Ultrafast
  pseudospin dynamics in graphene},}\ }\href@noop {} {\bibfield  {journal}
  {\bibinfo  {journal} {Phys. Rev. B}\ }\textbf {\bibinfo {volume} {92}},\
  \bibinfo {pages} {165429} (\bibinfo {year} {2015})}\BibitemShut {NoStop}%
\bibitem [{\citenamefont {Motlagh}\ \emph
  {et~al.}(2018{\natexlab{b}})\citenamefont {Motlagh}, \citenamefont {Wu},
  \citenamefont {Apalkov},\ and\ \citenamefont
  {Stockman}}]{Stockman_et_al_PhysRevB.98_2018_Rapid_Communication_Topological_Resonances}%
  \BibitemOpen
  \bibfield  {author} {\bibinfo {author} {\bibfnamefont {S.~A.~Oliaei}\
  \bibnamefont {Motlagh}}, \bibinfo {author} {\bibfnamefont {J.-S.}\
  \bibnamefont {Wu}}, \bibinfo {author} {\bibfnamefont {V.}~\bibnamefont
  {Apalkov}}, \ and\ \bibinfo {author} {\bibfnamefont {M.~I.}\ \bibnamefont
  {Stockman}},\ }\bibfield  {title} {\enquote {\bibinfo {title} {Femtosecond
  valley polarization and topological resonances in transition metal
  dichalcogenides},}\ }\href@noop {} {\bibfield  {journal} {\bibinfo  {journal}
  {Phys. Rev. B}\ }\textbf {\bibinfo {volume} {98}},\ \bibinfo {pages}
  {081406(R)--1--6} (\bibinfo {year} {2018}{\natexlab{b}})}\BibitemShut
  {NoStop}%
\bibitem [{\citenamefont {Sun}\ \emph {et~al.}(2017)\citenamefont {Sun},
  \citenamefont {Lai}, \citenamefont {Ma}, \citenamefont {Wang},\ and\
  \citenamefont
  {Liu}}]{Sun_et_al_Chinese_Physics_B_2017_Ultrafast_pulses_TMDC}%
  \BibitemOpen
  \bibfield  {author} {\bibinfo {author} {\bibfnamefont {D.}~\bibnamefont
  {Sun}}, \bibinfo {author} {\bibfnamefont {J.~W.}\ \bibnamefont {Lai}},
  \bibinfo {author} {\bibfnamefont {J.~C.}\ \bibnamefont {Ma}}, \bibinfo
  {author} {\bibfnamefont {Q.~S.}\ \bibnamefont {Wang}}, \ and\ \bibinfo
  {author} {\bibfnamefont {J.}~\bibnamefont {Liu}},\ }\bibfield  {title}
  {\enquote {\bibinfo {title} {Review of ultrafast spectroscopy studies of
  valley carrier dynamics in two-dimensional semiconducting transition metal
  dichalcogenides},}\ }\href {\doibase 10.1088/1674-1056/26/3/037801}
  {\bibfield  {journal} {\bibinfo  {journal} {Chin. Phys. B}\ }\textbf
  {\bibinfo {volume} {26}} (\bibinfo {year} {2017}),\
  10.1088/1674-1056/26/3/037801}\BibitemShut {NoStop}%
\bibitem [{\citenamefont {Zhang}\ \emph {et~al.}(2018)\citenamefont {Zhang},
  \citenamefont {Ouyang}, \citenamefont {Zheng}, \citenamefont {You},
  \citenamefont {Chen}, \citenamefont {Zhou}, \citenamefont {Sui},
  \citenamefont {Liu}, \citenamefont {Cheng},\ and\ \citenamefont
  {Jiang}}]{Zhang_et_al_OSA_2018_ultrafast_pulse_TMDC}%
  \BibitemOpen
  \bibfield  {author} {\bibinfo {author} {\bibfnamefont {Jun}\ \bibnamefont
  {Zhang}}, \bibinfo {author} {\bibfnamefont {Hao}\ \bibnamefont {Ouyang}},
  \bibinfo {author} {\bibfnamefont {Xin}\ \bibnamefont {Zheng}}, \bibinfo
  {author} {\bibfnamefont {Jie}\ \bibnamefont {You}}, \bibinfo {author}
  {\bibfnamefont {Runze}\ \bibnamefont {Chen}}, \bibinfo {author}
  {\bibfnamefont {Tong}\ \bibnamefont {Zhou}}, \bibinfo {author} {\bibfnamefont
  {Yizhen}\ \bibnamefont {Sui}}, \bibinfo {author} {\bibfnamefont
  {Yu}~\bibnamefont {Liu}}, \bibinfo {author} {\bibfnamefont {Xiang'ai}\
  \bibnamefont {Cheng}}, \ and\ \bibinfo {author} {\bibfnamefont {Tian}\
  \bibnamefont {Jiang}},\ }\bibfield  {title} {\enquote {\bibinfo {title}
  {Ultrafast saturable absorption of mos2 nanosheets under different
  pulse-width excitation conditions},}\ }\href {\doibase 10.1364/OL.43.000243}
  {\bibfield  {journal} {\bibinfo  {journal} {Opt. Lett.}\ }\textbf {\bibinfo
  {volume} {43}},\ \bibinfo {pages} {243--246} (\bibinfo {year}
  {2018})}\BibitemShut {NoStop}%
\bibitem [{\citenamefont {You}\ \emph {et~al.}(2017)\citenamefont {You},
  \citenamefont {Yin}, \citenamefont {Wu}, \citenamefont {Chew}, \citenamefont
  {Ren}, \citenamefont {Zhuang}, \citenamefont {Gholam-Mirzaei}, \citenamefont
  {Chini}, \citenamefont {Chang},\ and\ \citenamefont
  {Ghimire}}]{Ghimire_et_al_Nature_Communications_2017_HHG}%
  \BibitemOpen
  \bibfield  {author} {\bibinfo {author} {\bibfnamefont {Yong~Sing}\
  \bibnamefont {You}}, \bibinfo {author} {\bibfnamefont {Yanchun}\ \bibnamefont
  {Yin}}, \bibinfo {author} {\bibfnamefont {Yi}~\bibnamefont {Wu}}, \bibinfo
  {author} {\bibfnamefont {Andrew}\ \bibnamefont {Chew}}, \bibinfo {author}
  {\bibfnamefont {Xiaoming}\ \bibnamefont {Ren}}, \bibinfo {author}
  {\bibfnamefont {Fengjiang}\ \bibnamefont {Zhuang}}, \bibinfo {author}
  {\bibfnamefont {Shima}\ \bibnamefont {Gholam-Mirzaei}}, \bibinfo {author}
  {\bibfnamefont {Michael}\ \bibnamefont {Chini}}, \bibinfo {author}
  {\bibfnamefont {Zenghu}\ \bibnamefont {Chang}}, \ and\ \bibinfo {author}
  {\bibfnamefont {Shambhu}\ \bibnamefont {Ghimire}},\ }\bibfield  {title}
  {\enquote {\bibinfo {title} {High-harmonic generation in amorphous solids},}\
  }\href {\doibase 10.1038/s41467-017-00989-4} {\bibfield  {journal} {\bibinfo
  {journal} {Nat. Commun.}\ }\textbf {\bibinfo {volume} {8}},\ \bibinfo {pages}
  {724} (\bibinfo {year} {2017})}\BibitemShut {NoStop}%
\bibitem [{\citenamefont {Liu}\ \emph {et~al.}(2017)\citenamefont {Liu},
  \citenamefont {Li}, \citenamefont {You}, \citenamefont {Ghimire},
  \citenamefont {Heinz},\ and\ \citenamefont
  {Reis}}]{Reis_et_al_Nat_Phys_2017_HHG_from_2D_Crystals}%
  \BibitemOpen
  \bibfield  {author} {\bibinfo {author} {\bibfnamefont {H.~Z.}\ \bibnamefont
  {Liu}}, \bibinfo {author} {\bibfnamefont {Y.~L.}\ \bibnamefont {Li}},
  \bibinfo {author} {\bibfnamefont {Y.~S.}\ \bibnamefont {You}}, \bibinfo
  {author} {\bibfnamefont {S.}~\bibnamefont {Ghimire}}, \bibinfo {author}
  {\bibfnamefont {T.~F.}\ \bibnamefont {Heinz}}, \ and\ \bibinfo {author}
  {\bibfnamefont {D.~A.}\ \bibnamefont {Reis}},\ }\bibfield  {title} {\enquote
  {\bibinfo {title} {High-harmonic generation from an atomically thin
  semiconductor},}\ }\href@noop {} {\bibfield  {journal} {\bibinfo  {journal}
  {Nat. Phys.}\ }\textbf {\bibinfo {volume} {13}},\ \bibinfo {pages} {262--266}
  (\bibinfo {year} {2017})}\BibitemShut {NoStop}%
\bibitem [{\citenamefont {Kaiser}\ \emph {et~al.}(2000)\citenamefont {Kaiser},
  \citenamefont {Rethfeld}, \citenamefont {Vicanek},\ and\ \citenamefont
  {Simon}}]{Simon_et_al_PRB_2000_Strong_Field_Fs_Ionization_of_Dielectrics}%
  \BibitemOpen
  \bibfield  {author} {\bibinfo {author} {\bibfnamefont {A.}~\bibnamefont
  {Kaiser}}, \bibinfo {author} {\bibfnamefont {B.}~\bibnamefont {Rethfeld}},
  \bibinfo {author} {\bibfnamefont {M.}~\bibnamefont {Vicanek}}, \ and\
  \bibinfo {author} {\bibfnamefont {G.}~\bibnamefont {Simon}},\ }\bibfield
  {title} {\enquote {\bibinfo {title} {Microscopic processes in dielectrics
  under irradiation by subpicosecond laser pulses},}\ }\href@noop {} {\bibfield
   {journal} {\bibinfo  {journal} {Phys. Rev. B}\ }\textbf {\bibinfo {volume}
  {61}},\ \bibinfo {pages} {11437--11450} (\bibinfo {year} {2000})}\BibitemShut
  {NoStop}%
\bibitem [{\citenamefont {Rosa}\ \emph {et~al.}(2017)\citenamefont {Rosa},
  \citenamefont {Castaneda}, \citenamefont {Cruz}, \citenamefont {Padilha},
  \citenamefont {Gomes}, \citenamefont {de~Souza},\ and\ \citenamefont
  {Fragnito}}]{Rosa_et_al_Optical_Materials_Express_2017_stacking_graphene_saturable_absorbers}%
  \BibitemOpen
  \bibfield  {author} {\bibinfo {author} {\bibfnamefont {H.~G.}\ \bibnamefont
  {Rosa}}, \bibinfo {author} {\bibfnamefont {J.~A.}\ \bibnamefont {Castaneda}},
  \bibinfo {author} {\bibfnamefont {C.~H.~B.}\ \bibnamefont {Cruz}}, \bibinfo
  {author} {\bibfnamefont {L.~A.}\ \bibnamefont {Padilha}}, \bibinfo {author}
  {\bibfnamefont {J.~C.~V.}\ \bibnamefont {Gomes}}, \bibinfo {author}
  {\bibfnamefont {E.~A.~T.}\ \bibnamefont {de~Souza}}, \ and\ \bibinfo {author}
  {\bibfnamefont {H.~L.}\ \bibnamefont {Fragnito}},\ }\bibfield  {title}
  {\enquote {\bibinfo {title} {Controlled stacking of graphene monolayer
  saturable absorbers for ultrashort pulse generation in erbium-doped fiber
  lasers},}\ }\href {\doibase 10.1364/Ome.7.002528} {\bibfield  {journal}
  {\bibinfo  {journal} {Opt. Mater. Express}\ }\textbf {\bibinfo {volume}
  {7}},\ \bibinfo {pages} {2528--2537} (\bibinfo {year} {2017})}\BibitemShut
  {NoStop}%
\bibitem [{\citenamefont {Kumar}\ \emph {et~al.}(2009)\citenamefont {Kumar},
  \citenamefont {Anija}, \citenamefont {Kamaraju}, \citenamefont {Vasu},
  \citenamefont {Subrahmanyam}, \citenamefont {Sood},\ and\ \citenamefont
  {Rao}}]{Kumar_et_al_APL_2009_saturable_absorption_graphene}%
  \BibitemOpen
  \bibfield  {author} {\bibinfo {author} {\bibfnamefont {S.}~\bibnamefont
  {Kumar}}, \bibinfo {author} {\bibfnamefont {M.}~\bibnamefont {Anija}},
  \bibinfo {author} {\bibfnamefont {N.}~\bibnamefont {Kamaraju}}, \bibinfo
  {author} {\bibfnamefont {K.~S.}\ \bibnamefont {Vasu}}, \bibinfo {author}
  {\bibfnamefont {K.~S.}\ \bibnamefont {Subrahmanyam}}, \bibinfo {author}
  {\bibfnamefont {A.~K.}\ \bibnamefont {Sood}}, \ and\ \bibinfo {author}
  {\bibfnamefont {C.~N.~R.}\ \bibnamefont {Rao}},\ }\bibfield  {title}
  {\enquote {\bibinfo {title} {Femtosecond carrier dynamics and saturable
  absorption in graphene suspensions},}\ }\href {\doibase 10.1063/1.3264964}
  {\bibfield  {journal} {\bibinfo  {journal} {Appl. Phys. Lett.}\ }\textbf
  {\bibinfo {volume} {95}} (\bibinfo {year} {2009}),\
  10.1063/1.3264964}\BibitemShut {NoStop}%
\bibitem [{\citenamefont
  {Gesuele}(2019)}]{Gesuele_Photonics_2019_Transient_Absorption}%
  \BibitemOpen
  \bibfield  {author} {\bibinfo {author} {\bibfnamefont {F.}~\bibnamefont
  {Gesuele}},\ }\bibfield  {title} {\enquote {\bibinfo {title} {Ultrafast
  hyperspectral transient absorption spectroscopy: Application to single layer
  graphene},}\ }\href {\doibase 10.3390/photonics6030095} {\bibfield  {journal}
  {\bibinfo  {journal} {Photonics}\ }\textbf {\bibinfo {volume} {6}} (\bibinfo
  {year} {2019}),\ 10.3390/photonics6030095}\BibitemShut {NoStop}%
\bibitem [{\citenamefont {Geim}\ and\ \citenamefont
  {Novoselov}(2007)}]{Geim_et_al_Nat_Mater_2007_The_rise_of_graphene}%
  \BibitemOpen
  \bibfield  {author} {\bibinfo {author} {\bibfnamefont {A.~K.}\ \bibnamefont
  {Geim}}\ and\ \bibinfo {author} {\bibfnamefont {K.~S.}\ \bibnamefont
  {Novoselov}},\ }\bibfield  {title} {\enquote {\bibinfo {title} {The rise of
  graphene},}\ }\href@noop {} {\bibfield  {journal} {\bibinfo  {journal} {Nat
  Mater}\ }\textbf {\bibinfo {volume} {6}},\ \bibinfo {pages} {183--191}
  (\bibinfo {year} {2007})}\BibitemShut {NoStop}%
\bibitem [{\citenamefont {Neto}\ \emph {et~al.}(2009)\citenamefont {Neto},
  \citenamefont {Guinea}, \citenamefont {Peres}, \citenamefont {Novoselov},\
  and\ \citenamefont {Geim}}]{Electronic_properties_graphene_RMP_2009}%
  \BibitemOpen
  \bibfield  {author} {\bibinfo {author} {\bibfnamefont {A.~H.~Castro}\
  \bibnamefont {Neto}}, \bibinfo {author} {\bibfnamefont {F.}~\bibnamefont
  {Guinea}}, \bibinfo {author} {\bibfnamefont {N.~M.~R.}\ \bibnamefont
  {Peres}}, \bibinfo {author} {\bibfnamefont {K.~S.}\ \bibnamefont
  {Novoselov}}, \ and\ \bibinfo {author} {\bibfnamefont {A.~K.}\ \bibnamefont
  {Geim}},\ }\bibfield  {title} {\enquote {\bibinfo {title} {The electronic
  properties of graphene},}\ }\href@noop {} {\bibfield  {journal} {\bibinfo
  {journal} {Rev. Mod. Phys.}\ }\textbf {\bibinfo {volume} {81}},\ \bibinfo
  {pages} {109--162} (\bibinfo {year} {2009})}\BibitemShut {NoStop}%
\bibitem [{\citenamefont {Butler}\ \emph {et~al.}(2013)\citenamefont {Butler},
  \citenamefont {Hollen}, \citenamefont {Cao}, \citenamefont {Cui},
  \citenamefont {Gupta}, \citenamefont {Gutierrez}, \citenamefont {Heinz},
  \citenamefont {Hong}, \citenamefont {Huang}, \citenamefont {Ismach},
  \citenamefont {Johnston-Halperin}, \citenamefont {Kuno}, \citenamefont
  {Plashnitsa}, \citenamefont {Robinson}, \citenamefont {Ruoff}, \citenamefont
  {Salahuddin}, \citenamefont {Shan}, \citenamefont {Shi}, \citenamefont
  {Spencer}, \citenamefont {Terrones}, \citenamefont {Windl},\ and\
  \citenamefont {Goldberger}}]{Butler_et_al_Acs_Nano_2013_2D_Beyond_Graphene}%
  \BibitemOpen
  \bibfield  {author} {\bibinfo {author} {\bibfnamefont {S.~Z.}\ \bibnamefont
  {Butler}}, \bibinfo {author} {\bibfnamefont {S.~M.}\ \bibnamefont {Hollen}},
  \bibinfo {author} {\bibfnamefont {L.~Y.}\ \bibnamefont {Cao}}, \bibinfo
  {author} {\bibfnamefont {Y.}~\bibnamefont {Cui}}, \bibinfo {author}
  {\bibfnamefont {J.~A.}\ \bibnamefont {Gupta}}, \bibinfo {author}
  {\bibfnamefont {H.~R.}\ \bibnamefont {Gutierrez}}, \bibinfo {author}
  {\bibfnamefont {T.~F.}\ \bibnamefont {Heinz}}, \bibinfo {author}
  {\bibfnamefont {S.~S.}\ \bibnamefont {Hong}}, \bibinfo {author}
  {\bibfnamefont {J.~X.}\ \bibnamefont {Huang}}, \bibinfo {author}
  {\bibfnamefont {A.~F.}\ \bibnamefont {Ismach}}, \bibinfo {author}
  {\bibfnamefont {E.}~\bibnamefont {Johnston-Halperin}}, \bibinfo {author}
  {\bibfnamefont {M.}~\bibnamefont {Kuno}}, \bibinfo {author} {\bibfnamefont
  {V.~V.}\ \bibnamefont {Plashnitsa}}, \bibinfo {author} {\bibfnamefont
  {R.~D.}\ \bibnamefont {Robinson}}, \bibinfo {author} {\bibfnamefont {R.~S.}\
  \bibnamefont {Ruoff}}, \bibinfo {author} {\bibfnamefont {S.}~\bibnamefont
  {Salahuddin}}, \bibinfo {author} {\bibfnamefont {J.}~\bibnamefont {Shan}},
  \bibinfo {author} {\bibfnamefont {L.}~\bibnamefont {Shi}}, \bibinfo {author}
  {\bibfnamefont {M.~G.}\ \bibnamefont {Spencer}}, \bibinfo {author}
  {\bibfnamefont {M.}~\bibnamefont {Terrones}}, \bibinfo {author}
  {\bibfnamefont {W.}~\bibnamefont {Windl}}, \ and\ \bibinfo {author}
  {\bibfnamefont {J.~E.}\ \bibnamefont {Goldberger}},\ }\bibfield  {title}
  {\enquote {\bibinfo {title} {Progress, challenges, and opportunities in
  two-dimensional materials beyond graphene},}\ }\href {\doibase
  10.1021/nn400280c} {\bibfield  {journal} {\bibinfo  {journal} {Acs Nano}\
  }\textbf {\bibinfo {volume} {7}},\ \bibinfo {pages} {2898--2926} (\bibinfo
  {year} {2013})}\BibitemShut {NoStop}%
\bibitem [{\citenamefont {Novoselov}\ \emph {et~al.}(2005)\citenamefont
  {Novoselov}, \citenamefont {Geim}, \citenamefont {Morozov}, \citenamefont
  {Jiang}, \citenamefont {Katsnelson}, \citenamefont {Grigorieva},
  \citenamefont {Dubonos},\ and\ \citenamefont
  {Firsov}}]{Novoselov_Geim_et_al_nature04233_2D_Electrons_in_Graphene}%
  \BibitemOpen
  \bibfield  {author} {\bibinfo {author} {\bibfnamefont {K.~S.}\ \bibnamefont
  {Novoselov}}, \bibinfo {author} {\bibfnamefont {A.~K.}\ \bibnamefont {Geim}},
  \bibinfo {author} {\bibfnamefont {S.~V.}\ \bibnamefont {Morozov}}, \bibinfo
  {author} {\bibfnamefont {D.}~\bibnamefont {Jiang}}, \bibinfo {author}
  {\bibfnamefont {M.~I.}\ \bibnamefont {Katsnelson}}, \bibinfo {author}
  {\bibfnamefont {I.~V.}\ \bibnamefont {Grigorieva}}, \bibinfo {author}
  {\bibfnamefont {S.~V.}\ \bibnamefont {Dubonos}}, \ and\ \bibinfo {author}
  {\bibfnamefont {A.~A.}\ \bibnamefont {Firsov}},\ }\bibfield  {title}
  {\enquote {\bibinfo {title} {Two-dimensional gas of massless {Dirac} fermions
  in graphene},}\ }\href@noop {} {\bibfield  {journal} {\bibinfo  {journal}
  {Nature}\ }\textbf {\bibinfo {volume} {438}},\ \bibinfo {pages} {197--200}
  (\bibinfo {year} {2005})}\BibitemShut {NoStop}%
\bibitem [{\citenamefont {Kormanyos}\ \emph {et~al.}(2015)\citenamefont
  {Kormanyos}, \citenamefont {Burkard}, \citenamefont {Gmitra}, \citenamefont
  {Fabian}, \citenamefont {Zolyomi}, \citenamefont {Drummond},\ and\
  \citenamefont
  {Fal'ko}}]{Kormanyos_et_al_2d_Materials_2015_k.p_theory_for_two_dimensional}%
  \BibitemOpen
  \bibfield  {author} {\bibinfo {author} {\bibfnamefont {A.}~\bibnamefont
  {Kormanyos}}, \bibinfo {author} {\bibfnamefont {G.}~\bibnamefont {Burkard}},
  \bibinfo {author} {\bibfnamefont {M.}~\bibnamefont {Gmitra}}, \bibinfo
  {author} {\bibfnamefont {J.}~\bibnamefont {Fabian}}, \bibinfo {author}
  {\bibfnamefont {V.}~\bibnamefont {Zolyomi}}, \bibinfo {author} {\bibfnamefont
  {N.~D.}\ \bibnamefont {Drummond}}, \ and\ \bibinfo {author} {\bibfnamefont
  {V.}~\bibnamefont {Fal'ko}},\ }\bibfield  {title} {\enquote {\bibinfo {title}
  {k.p theory for two-dimensional transition metal dichalcogenide
  semiconductors (vol 2, 022001, 2015)},}\ }\href@noop {} {\bibfield  {journal}
  {\bibinfo  {journal} {2d Materials}\ }\textbf {\bibinfo {volume} {2}}
  (\bibinfo {year} {2015})}\BibitemShut {NoStop}%
\bibitem [{\citenamefont {Ye}\ \emph {et~al.}(2016)\citenamefont {Ye},
  \citenamefont {Xiao}, \citenamefont {Wang}, \citenamefont {Ye}, \citenamefont
  {Zhu}, \citenamefont {Zhao}, \citenamefont {Wang}, \citenamefont {Zhao},
  \citenamefont {Yin},\ and\ \citenamefont
  {Zhang}}]{Ye_et_al_Nature_Nanotechnology_2016_Electrical_generation_and_control}%
  \BibitemOpen
  \bibfield  {author} {\bibinfo {author} {\bibfnamefont {Y.}~\bibnamefont
  {Ye}}, \bibinfo {author} {\bibfnamefont {J.}~\bibnamefont {Xiao}}, \bibinfo
  {author} {\bibfnamefont {H.~L.}\ \bibnamefont {Wang}}, \bibinfo {author}
  {\bibfnamefont {Z.~L.}\ \bibnamefont {Ye}}, \bibinfo {author} {\bibfnamefont
  {H.~Y.}\ \bibnamefont {Zhu}}, \bibinfo {author} {\bibfnamefont
  {M.}~\bibnamefont {Zhao}}, \bibinfo {author} {\bibfnamefont {Y.}~\bibnamefont
  {Wang}}, \bibinfo {author} {\bibfnamefont {J.~H.}\ \bibnamefont {Zhao}},
  \bibinfo {author} {\bibfnamefont {X.~B.}\ \bibnamefont {Yin}}, \ and\
  \bibinfo {author} {\bibfnamefont {X.}~\bibnamefont {Zhang}},\ }\bibfield
  {title} {\enquote {\bibinfo {title} {Electrical generation and control of the
  valley carriers in a monolayer transition metal dichalcogenide},}\
  }\href@noop {} {\bibfield  {journal} {\bibinfo  {journal} {Nat.
  Nanotechnol.}\ }\textbf {\bibinfo {volume} {11}},\ \bibinfo {pages}
  {598--602} (\bibinfo {year} {2016})}\BibitemShut {NoStop}%
\bibitem [{\citenamefont {Jariwala}\ \emph {et~al.}(2014)\citenamefont
  {Jariwala}, \citenamefont {Sangwan}, \citenamefont {Lauhon}, \citenamefont
  {Marks},\ and\ \citenamefont
  {Hersam}}]{Jariwala_et_al_Asc_Nano_2014_Transition_Metal}%
  \BibitemOpen
  \bibfield  {author} {\bibinfo {author} {\bibfnamefont {D.}~\bibnamefont
  {Jariwala}}, \bibinfo {author} {\bibfnamefont {V.~K.}\ \bibnamefont
  {Sangwan}}, \bibinfo {author} {\bibfnamefont {L.~J.}\ \bibnamefont {Lauhon}},
  \bibinfo {author} {\bibfnamefont {T.~J.}\ \bibnamefont {Marks}}, \ and\
  \bibinfo {author} {\bibfnamefont {M.~C.}\ \bibnamefont {Hersam}},\ }\bibfield
   {title} {\enquote {\bibinfo {title} {Emerging device applications for
  semiconducting two-dimensional transition metal dichalcogenides},}\
  }\href@noop {} {\bibfield  {journal} {\bibinfo  {journal} {Acs Nano}\
  }\textbf {\bibinfo {volume} {8}},\ \bibinfo {pages} {1102--1120} (\bibinfo
  {year} {2014})}\BibitemShut {NoStop}%
\bibitem [{\citenamefont {Oliaei~Motlagh}\ \emph {et~al.}(2019)\citenamefont
  {Oliaei~Motlagh}, \citenamefont {Nematollahi}, \citenamefont {Apalkov},\ and\
  \citenamefont {Stockman}}]{Stockman_et_al_PRB_2019_gapped_graphene}%
  \BibitemOpen
  \bibfield  {author} {\bibinfo {author} {\bibfnamefont {S.~Azar}\ \bibnamefont
  {Oliaei~Motlagh}}, \bibinfo {author} {\bibfnamefont {Fatemeh}\ \bibnamefont
  {Nematollahi}}, \bibinfo {author} {\bibfnamefont {Vadym}\ \bibnamefont
  {Apalkov}}, \ and\ \bibinfo {author} {\bibfnamefont {Mark~I.}\ \bibnamefont
  {Stockman}},\ }\bibfield  {title} {\enquote {\bibinfo {title} {Topological
  resonance and single-optical-cycle valley polarization in gapped graphene},}\
  }\href {\doibase 10.1103/PhysRevB.100.115431} {\bibfield  {journal} {\bibinfo
   {journal} {Phys. Rev. B}\ }\textbf {\bibinfo {volume} {100}},\ \bibinfo
  {pages} {115431} (\bibinfo {year} {2019})}\BibitemShut {NoStop}%
\bibitem [{\citenamefont {Motlagh}\ \emph
  {et~al.}(2019{\natexlab{a}})\citenamefont {Motlagh}, \citenamefont
  {Nematollahi}, \citenamefont {Mitra}, \citenamefont {Zafar}, \citenamefont
  {Apalkov},\ and\ \citenamefont
  {Stockman}}]{Stockman_et_al_J.Phys.Condens.Matter_2019_current_gapped_graphene}%
  \BibitemOpen
  \bibfield  {author} {\bibinfo {author} {\bibfnamefont {Seyyedeh Azar~Oliaei}\
  \bibnamefont {Motlagh}}, \bibinfo {author} {\bibfnamefont {Fatemeh}\
  \bibnamefont {Nematollahi}}, \bibinfo {author} {\bibfnamefont {Aranyo}\
  \bibnamefont {Mitra}}, \bibinfo {author} {\bibfnamefont {Ahmal~Jawad}\
  \bibnamefont {Zafar}}, \bibinfo {author} {\bibfnamefont {Vadym}\ \bibnamefont
  {Apalkov}}, \ and\ \bibinfo {author} {\bibfnamefont {Mark~I}\ \bibnamefont
  {Stockman}},\ }\bibfield  {title} {\enquote {\bibinfo {title} {Ultrafast
  optical},}\ }\href {\doibase 10.1088/1361-648X/ab4fc7} {\bibfield  {journal}
  {\bibinfo  {journal} {Journal of Physics: Condensed Matter}\ } (\bibinfo
  {year} {2019}{\natexlab{a}}),\ 10.1088/1361-648X/ab4fc7}\BibitemShut
  {NoStop}%
\bibitem [{\citenamefont {Hwang}\ and\ \citenamefont
  {Sarma}(2008)}]{Hwang_Das_Sarma_PRB_2008_Graphene_Relaxation_Time}%
  \BibitemOpen
  \bibfield  {author} {\bibinfo {author} {\bibfnamefont {E.~H.}\ \bibnamefont
  {Hwang}}\ and\ \bibinfo {author} {\bibfnamefont {S.~Das}\ \bibnamefont
  {Sarma}},\ }\bibfield  {title} {\enquote {\bibinfo {title} {Single-particle
  relaxation time versus transport scattering time in a two-dimensional
  graphene layer},}\ }\href@noop {} {\bibfield  {journal} {\bibinfo  {journal}
  {Phys. Rev. B}\ }\textbf {\bibinfo {volume} {77}},\ \bibinfo {pages}
  {195412--1--6} (\bibinfo {year} {2008})}\BibitemShut {NoStop}%
\bibitem [{\citenamefont {Breusing}\ \emph {et~al.}(2011)\citenamefont
  {Breusing}, \citenamefont {Kuehn}, \citenamefont {Winzer}, \citenamefont
  {Malic}, \citenamefont {Milde}, \citenamefont {Severin}, \citenamefont
  {Rabe}, \citenamefont {Ropers}, \citenamefont {Knorr},\ and\ \citenamefont
  {Elsaesser}}]{Breusing_et_al_Ultrafast-nonequilibrium-carrier-dynamics_PRB_2011}%
  \BibitemOpen
  \bibfield  {author} {\bibinfo {author} {\bibfnamefont {M.}~\bibnamefont
  {Breusing}}, \bibinfo {author} {\bibfnamefont {S.}~\bibnamefont {Kuehn}},
  \bibinfo {author} {\bibfnamefont {T.}~\bibnamefont {Winzer}}, \bibinfo
  {author} {\bibfnamefont {E.}~\bibnamefont {Malic}}, \bibinfo {author}
  {\bibfnamefont {F.}~\bibnamefont {Milde}}, \bibinfo {author} {\bibfnamefont
  {N.}~\bibnamefont {Severin}}, \bibinfo {author} {\bibfnamefont {J.~P.}\
  \bibnamefont {Rabe}}, \bibinfo {author} {\bibfnamefont {C.}~\bibnamefont
  {Ropers}}, \bibinfo {author} {\bibfnamefont {A.}~\bibnamefont {Knorr}}, \
  and\ \bibinfo {author} {\bibfnamefont {T.}~\bibnamefont {Elsaesser}},\
  }\bibfield  {title} {\enquote {\bibinfo {title} {Ultrafast nonequilibrium
  carrier dynamics in a single graphene layer},}\ }\href
  {http://link.aps.org/doi/10.1103/PhysRevB.83.153410} {\bibfield  {journal}
  {\bibinfo  {journal} {Phys. Rev. B}\ }\textbf {\bibinfo {volume} {83}},\
  \bibinfo {pages} {153410} (\bibinfo {year} {2011})}\BibitemShut {NoStop}%
\bibitem [{\citenamefont {Malic}\ \emph {et~al.}(2011)\citenamefont {Malic},
  \citenamefont {Winzer}, \citenamefont {Bobkin},\ and\ \citenamefont
  {Knorr}}]{theory_absorption_ultrafast_kinetics_graphene_PRB_2011}%
  \BibitemOpen
  \bibfield  {author} {\bibinfo {author} {\bibfnamefont {Ermin}\ \bibnamefont
  {Malic}}, \bibinfo {author} {\bibfnamefont {Torben}\ \bibnamefont {Winzer}},
  \bibinfo {author} {\bibfnamefont {Evgeny}\ \bibnamefont {Bobkin}}, \ and\
  \bibinfo {author} {\bibfnamefont {Andreas}\ \bibnamefont {Knorr}},\
  }\bibfield  {title} {\enquote {\bibinfo {title} {Microscopic theory of
  absorption and ultrafast many-particle kinetics in graphene},}\ }\href@noop
  {} {\bibfield  {journal} {\bibinfo  {journal} {Phys. Rev. B}\ }\textbf
  {\bibinfo {volume} {84}},\ \bibinfo {pages} {205406} (\bibinfo {year}
  {2011})}\BibitemShut {NoStop}%
\bibitem [{\citenamefont {Brida}\ \emph {et~al.}(2013)\citenamefont {Brida},
  \citenamefont {Tomadin}, \citenamefont {Manzoni}, \citenamefont {Kim},
  \citenamefont {Lombardo}, \citenamefont {Milana}, \citenamefont {Nair},
  \citenamefont {Novoselov}, \citenamefont {Ferrari}, \citenamefont {Cerullo},\
  and\ \citenamefont
  {Polini}}]{Ultrafast_collinear_scattering_graphene_nat_comm_2013}%
  \BibitemOpen
  \bibfield  {author} {\bibinfo {author} {\bibfnamefont {D.}~\bibnamefont
  {Brida}}, \bibinfo {author} {\bibfnamefont {A.}~\bibnamefont {Tomadin}},
  \bibinfo {author} {\bibfnamefont {C.}~\bibnamefont {Manzoni}}, \bibinfo
  {author} {\bibfnamefont {Y.~J.}\ \bibnamefont {Kim}}, \bibinfo {author}
  {\bibfnamefont {A.}~\bibnamefont {Lombardo}}, \bibinfo {author}
  {\bibfnamefont {S.}~\bibnamefont {Milana}}, \bibinfo {author} {\bibfnamefont
  {R.~R.}\ \bibnamefont {Nair}}, \bibinfo {author} {\bibfnamefont {K.~S.}\
  \bibnamefont {Novoselov}}, \bibinfo {author} {\bibfnamefont {A.~C.}\
  \bibnamefont {Ferrari}}, \bibinfo {author} {\bibfnamefont {G.}~\bibnamefont
  {Cerullo}}, \ and\ \bibinfo {author} {\bibfnamefont {M.}~\bibnamefont
  {Polini}},\ }\bibfield  {title} {\enquote {\bibinfo {title} {Ultrafast
  collinear scattering and carrier multiplication in graphene},}\ }\href
  {\doibase 10.1038/ncomms2987} {\bibfield  {journal} {\bibinfo  {journal} {Nat
  Commun}\ }\textbf {\bibinfo {volume} {4}},\ \bibinfo {pages} {1987--1--9}
  (\bibinfo {year} {2013})}\BibitemShut {NoStop}%
\bibitem [{\citenamefont {Gierz}\ \emph {et~al.}(2013)\citenamefont {Gierz},
  \citenamefont {Petersen}, \citenamefont {Mitrano}, \citenamefont {Cacho},
  \citenamefont {Turcu}, \citenamefont {Springate}, \citenamefont {Stohr},
  \citenamefont {Kohler}, \citenamefont {Starke},\ and\ \citenamefont
  {Cavalleri}}]{Gierz_Snapshots-non-equilibrium-Dirac_Nat-Material_2013}%
  \BibitemOpen
  \bibfield  {author} {\bibinfo {author} {\bibfnamefont {I.}~\bibnamefont
  {Gierz}}, \bibinfo {author} {\bibfnamefont {J.~C.}\ \bibnamefont {Petersen}},
  \bibinfo {author} {\bibfnamefont {M.}~\bibnamefont {Mitrano}}, \bibinfo
  {author} {\bibfnamefont {C.}~\bibnamefont {Cacho}}, \bibinfo {author}
  {\bibfnamefont {I.~C.}\ \bibnamefont {Turcu}}, \bibinfo {author}
  {\bibfnamefont {E.}~\bibnamefont {Springate}}, \bibinfo {author}
  {\bibfnamefont {A.}~\bibnamefont {Stohr}}, \bibinfo {author} {\bibfnamefont
  {A.}~\bibnamefont {Kohler}}, \bibinfo {author} {\bibfnamefont
  {U.}~\bibnamefont {Starke}}, \ and\ \bibinfo {author} {\bibfnamefont
  {A.}~\bibnamefont {Cavalleri}},\ }\bibfield  {title} {\enquote {\bibinfo
  {title} {Snapshots of non-equilibrium {Dirac} carrier distributions in
  graphene},}\ }\href {\doibase 10.1038/nmat3757} {\bibfield  {journal}
  {\bibinfo  {journal} {Nat. Mater.}\ }\textbf {\bibinfo {volume} {12}},\
  \bibinfo {pages} {1119--24} (\bibinfo {year} {2013})}\BibitemShut {NoStop}%
\bibitem [{\citenamefont {Tomadin}\ \emph {et~al.}(2013)\citenamefont
  {Tomadin}, \citenamefont {Brida}, \citenamefont {Cerullo}, \citenamefont
  {Ferrari},\ and\ \citenamefont
  {Polini}}]{Nonequilibrium_dynamics_photoexcited_electrons_graphene_PRB_2013}%
  \BibitemOpen
  \bibfield  {author} {\bibinfo {author} {\bibfnamefont {Andrea}\ \bibnamefont
  {Tomadin}}, \bibinfo {author} {\bibfnamefont {Daniele}\ \bibnamefont
  {Brida}}, \bibinfo {author} {\bibfnamefont {Giulio}\ \bibnamefont {Cerullo}},
  \bibinfo {author} {\bibfnamefont {Andrea~C.}\ \bibnamefont {Ferrari}}, \ and\
  \bibinfo {author} {\bibfnamefont {Marco}\ \bibnamefont {Polini}},\ }\bibfield
   {title} {\enquote {\bibinfo {title} {Nonequilibrium dynamics of photoexcited
  electrons in graphene: Collinear scattering, {Auger} processes, and the
  impact of screening},}\ }\href@noop {} {\bibfield  {journal} {\bibinfo
  {journal} {Phys. Rev. B}\ }\textbf {\bibinfo {volume} {88}},\ \bibinfo
  {pages} {035430} (\bibinfo {year} {2013})}\BibitemShut {NoStop}%
\bibitem [{\citenamefont {Zhou}\ \emph {et~al.}(2007)\citenamefont {Zhou},
  \citenamefont {Gweon}, \citenamefont {Fedorov}, \citenamefont {First},
  \citenamefont {de~Heer}, \citenamefont {Lee}, \citenamefont {Guinea},
  \citenamefont {Neto},\ and\ \citenamefont
  {Lanzara}}]{Lanzara_et_al_Nat_Mat_2007_Gapped_Graphene}%
  \BibitemOpen
  \bibfield  {author} {\bibinfo {author} {\bibfnamefont {S.~Y.}\ \bibnamefont
  {Zhou}}, \bibinfo {author} {\bibfnamefont {G.~H.}\ \bibnamefont {Gweon}},
  \bibinfo {author} {\bibfnamefont {A.~V.}\ \bibnamefont {Fedorov}}, \bibinfo
  {author} {\bibfnamefont {P.~N.}\ \bibnamefont {First}}, \bibinfo {author}
  {\bibfnamefont {W.~A.}\ \bibnamefont {de~Heer}}, \bibinfo {author}
  {\bibfnamefont {D.~H.}\ \bibnamefont {Lee}}, \bibinfo {author} {\bibfnamefont
  {F.}~\bibnamefont {Guinea}}, \bibinfo {author} {\bibfnamefont {A.~H.~Castro}\
  \bibnamefont {Neto}}, \ and\ \bibinfo {author} {\bibfnamefont
  {A.}~\bibnamefont {Lanzara}},\ }\bibfield  {title} {\enquote {\bibinfo
  {title} {Substrate-induced bandgap opening in epitaxial graphene},}\
  }\href@noop {} {\bibfield  {journal} {\bibinfo  {journal} {Nat. Mater.}\
  }\textbf {\bibinfo {volume} {6}},\ \bibinfo {pages} {770} (\bibinfo {year}
  {2007})}\BibitemShut {NoStop}%
\bibitem [{\citenamefont {Pedersen}\ \emph {et~al.}(2009)\citenamefont
  {Pedersen}, \citenamefont {Jauho},\ and\ \citenamefont
  {Pedersen}}]{Kjeld_et_al_PhysRevB.79.113406_2009_Gapped_Graphene_Optical_Response}%
  \BibitemOpen
  \bibfield  {author} {\bibinfo {author} {\bibfnamefont {Thomas~G.}\
  \bibnamefont {Pedersen}}, \bibinfo {author} {\bibfnamefont {Antti-Pekka}\
  \bibnamefont {Jauho}}, \ and\ \bibinfo {author} {\bibfnamefont {Kjeld}\
  \bibnamefont {Pedersen}},\ }\bibfield  {title} {\enquote {\bibinfo {title}
  {Optical response and excitons in gapped graphene},}\ }\href {\doibase
  10.1103/PhysRevB.79.113406} {\bibfield  {journal} {\bibinfo  {journal} {Phys.
  Rev. B}\ }\textbf {\bibinfo {volume} {79}},\ \bibinfo {pages} {113406}
  (\bibinfo {year} {2009})}\BibitemShut {NoStop}%
\bibitem [{\citenamefont
  {Pyatkovskiy}(2008)}]{Pyatkovskiy_JPCM_2008_Plasmons_in_Gapped_Graqphene}%
  \BibitemOpen
  \bibfield  {author} {\bibinfo {author} {\bibfnamefont {P.~K.}\ \bibnamefont
  {Pyatkovskiy}},\ }\bibfield  {title} {\enquote {\bibinfo {title} {Dynamical
  polarization, screening, and plasmons in gapped graphene},}\ }\href@noop {}
  {\bibfield  {journal} {\bibinfo  {journal} {J. Condens. Matter Phys.}\
  }\textbf {\bibinfo {volume} {21}},\ \bibinfo {pages} {025506} (\bibinfo
  {year} {2008})}\BibitemShut {NoStop}%
\bibitem [{\citenamefont
  {Bloch}(1929)}]{Bloch_Z_Phys_1929_Functions_Oscillations_in_Crystals}%
  \BibitemOpen
  \bibfield  {author} {\bibinfo {author} {\bibfnamefont {F.}~\bibnamefont
  {Bloch}},\ }\bibfield  {title} {\enquote {\bibinfo {title} {{\"Uber} die
  {Quantenmechanik} der {Elektronen} in {Kristallgittern}},}\ }\href@noop {}
  {\bibfield  {journal} {\bibinfo  {journal} {Z. Phys. A}\ }\textbf {\bibinfo
  {volume} {52}},\ \bibinfo {pages} {555--600} (\bibinfo {year}
  {1929})}\BibitemShut {NoStop}%
\bibitem [{\citenamefont {Kelardeh}\ \emph {et~al.}(2016)\citenamefont
  {Kelardeh}, \citenamefont {Apalkov},\ and\ \citenamefont
  {Stockman}}]{Stockman_et_al_PhysRevB.93.155434_Graphene_Circular_Interferometry}%
  \BibitemOpen
  \bibfield  {author} {\bibinfo {author} {\bibfnamefont {H.~K.}\ \bibnamefont
  {Kelardeh}}, \bibinfo {author} {\bibfnamefont {V.}~\bibnamefont {Apalkov}}, \
  and\ \bibinfo {author} {\bibfnamefont {M.~I.}\ \bibnamefont {Stockman}},\
  }\bibfield  {title} {\enquote {\bibinfo {title} {Attosecond strong-field
  interferometry in graphene: Chirality, singularity, and {Berry} phase},}\
  }\href@noop {} {\bibfield  {journal} {\bibinfo  {journal} {Phys. Rev. B}\
  }\textbf {\bibinfo {volume} {93}},\ \bibinfo {pages} {155434--1--7} (\bibinfo
  {year} {2016})}\BibitemShut {NoStop}%
\bibitem [{\citenamefont
  {Houston}(1940)}]{Houston_PR_1940_Electron_Acceleration_in_Lattice}%
  \BibitemOpen
  \bibfield  {author} {\bibinfo {author} {\bibfnamefont {W.~V.}\ \bibnamefont
  {Houston}},\ }\bibfield  {title} {\enquote {\bibinfo {title} {Acceleration of
  electrons in a crystal lattice},}\ }\href@noop {} {\bibfield  {journal}
  {\bibinfo  {journal} {Phys. Rev.}\ }\textbf {\bibinfo {volume} {57}},\
  \bibinfo {pages} {184--186} (\bibinfo {year} {1940})}\BibitemShut {NoStop}%
\bibitem [{\citenamefont {Motlagh}\ \emph
  {et~al.}(2019{\natexlab{b}})\citenamefont {Motlagh}, \citenamefont
  {Nematollahi}, \citenamefont {Apalkov},\ and\ \citenamefont
  {Stockman}}]{Stockman_et_al_PhysRevB.100.115431_2019_Gapped_Graphene}%
  \BibitemOpen
  \bibfield  {author} {\bibinfo {author} {\bibfnamefont {S.~A.~Oliaei}\
  \bibnamefont {Motlagh}}, \bibinfo {author} {\bibfnamefont {F.}~\bibnamefont
  {Nematollahi}}, \bibinfo {author} {\bibfnamefont {V.}~\bibnamefont
  {Apalkov}}, \ and\ \bibinfo {author} {\bibfnamefont {M.~I.}\ \bibnamefont
  {Stockman}},\ }\bibfield  {title} {\enquote {\bibinfo {title} {Topological
  resonance and single-optical-cycle valley polarization in gapped graphene},}\
  }\href@noop {} {\bibfield  {journal} {\bibinfo  {journal} {Phys. Rev. B}\
  }\textbf {\bibinfo {volume} {100}},\ \bibinfo {pages} {115431} (\bibinfo
  {year} {2019}{\natexlab{b}})}\BibitemShut {NoStop}%
\end{thebibliography}

\end{document}